\documentclass[a4paper,fleqn,usenatbib]{mnras}

\usepackage[T1]{fontenc}
\usepackage{ae,aecompl}

\usepackage{graphicx}	
\usepackage{amsmath}	
\usepackage{amssymb}	

\title[Constraining the Contribution of Galaxies and AGNs to Cosmic Reionization]{Constraining the Contribution of Galaxies and Active Galactic Nuclei to Cosmic Reionization}
\author[S. Yoshiura et al.]{
 Shintaro Yoshiura$^1$\thanks{E-mail: 161d9002@st.kumamoto-u.ac.jp},
 Kenji Hasegawa$^2$,
 Kiyotomo Ichiki$^{2,3}$, 
 Hiroyuki Tashiro$^2$,
\newauthor 
 Hayato Shimabukuro$^{4}$,
 Keitaro Takahashi$^1$
\\
$^{1}$Department of Physics, Kumamoto University, Kumamoto,Japan\\
$^{2}$Department of Physics, Nagoya University, Aichi , Japan\\
$^{3}$Kobayashi-Maskawa Institute for the Origin of Particles and the Universe, Nagoya University Nagoya, 464-8602, Japan\\
$^{4}$Sorbonne Universites, UPMC, LERMA, Observatoire de Paris, PSL research university, CNRS, F-75014, Paris, France
}

\date{Accepted XXX. Received YYY; in original form ZZZ}

\pubyear{2015}

\begin{document}
\label{firstpage}
\pagerange{\pageref{firstpage}--\pageref{lastpage}}
\maketitle

\begin{abstract}
Understanding the detailed process of cosmic reionization is one of the remaining problems in astrophysics and cosmology.  Here we construct a model of cosmic reionization that includes contribution from high-$z$ galaxies and active galactic nuclei (AGNs), and calculate the reionization and thermal histories with the model. 
To keep the model to be general and realistic, we vary the escape fraction of ionizing photons, $f_{\rm esc}$, and the faint-end slope of the AGN luminosity function at high redshifts, $\alpha_{\rm hz}$, within the constraints from the observed cosmic star formation history and the observed bright-end UV luminosity functions at $z \leq 6$. Additionally, we model the spectral energy distribution (SED) of AGNs which depends on the Eddington ratio and the black-hole mass. 
By comparing the computed reionization histories with the observed \ion{H}{i} fractions and the optical depth for Thomson scattering from Planck, we find that $\alpha_{\rm hz} > -1.5$ and $f_{\rm esc} < 0.15$ are favored when we employ the bright-end luminosity function obtained by Giallongo et al. 
Our result suggests that an AGN-dominated model with the abundance of faint-AGNs being as large as the estimate by Giallongo et al. is allowed only if the contribution from high-z galaxies is almost negligible, while a galaxy-dominant model is also allowed. 
We also find that the shape of SED has a significant impact on the thermal history. 
Therefore it is expected that measurements of the thermal state of the IGM provide useful information on properties of ionizing sources. 
\end{abstract}




\section{Introduction}\label{intro}
It is generally believed that the Universe became neutral at $z\approx 1100$ (cosmic recombination) and was gradually ionized again by ionizing sources formed following cosmic recombination. 
The time evolution of this global transition called cosmic reionization has been widely explored. 
One of observational approaches to estimate neutral fractions at high redshifts is to measure absorption features imprinted on high-$z$ quasar (QSO) spectra. 
The \ion{H}{i} Lyman-$\alpha$~(Ly$\alpha$) absorption lines imprinted on QSOs spectra have indicated that hydrogen was almost reionized by $z\sim6$ \citep{2006AJ....132..117F}. 
Similar observations measuring \ion{He}{ii} absorption lines have also shown that helium reionization was mostly completed by $z\approx2.7$ \citep{2011MNRAS.410.1096B, 2014arXiv1405.7405W,2014ApJ...784...42S}. 
Observing the redshift evolution of Ly$\alpha$
luminosity function ($\rm Ly{\alpha}$ LF) can also  provide valuable information on neutral hydrogen fractions during the epoch of reionization (EoR), because the detectability of $\rm Ly{\alpha}$ radiation is very sensitive to the amount of neutral hydrogen. 
Recent observations for $z\geq5.7$ Ly$\alpha$ emitters (LAEs) have shown
that the number density of LAEs decreases with redshift \citep{Ota10,
Hu10, 2010ApJ...723..869O, Kashikawa11, 2014ApJ...797...16K}.
This fact implies that the neutral hydrogen fraction increases from $z\sim6$ to $z\sim7$ \citep{2015MNRAS.452..261C}.
Furthermore, the measurements of the Cosmic Microwave Background (CMB) optical depth to Thomson scattering have imposed an additional constraint on the reionization history. 
The most recent observation by {\it Planck} has shown that the optical
depth $\tau_{\rm e}$ is $0.058\pm0.012$ which corresponds to an
instantaneous reionization redshift of $z_{\rm r}=8.5\pm{0.9}$ \citep{2016arXiv160503507P}. 
 
Despite the advancing understanding on the history of cosmic reionization, there is no crucial observation for the type of ionizing sources which was responsible for cosmic reionization. 
Revealing the type of ionizing sources is also important for understanding the process of reionization, since the mean free path of ionizing photons is essential for the reionization evolution. 

High-$z$ galaxies are thought to be plausible candidates for the ionizing sources, since a number of galaxies have already discovered at $z>6$ \citep{2009ApJ...706.1136O,2014ApJ...786..108O, 2015ApJ...803...34B}. 
In addition, some of recent observations have presented that the faint-end slopes of ultraviolet luminosity functions (UVLFs) during the EoR are steep and extended down to an absolute UV magnitude of $M_{\rm UV} \sim -16$ \citep{Atek15, 2015ApJ...799...12I}. 
These facts strengthen the possibility that the high-$z$ galaxies were main ionizing sources of reionization. 
\cite{Robertson15} have indeed shown that the Thomson scattering optical depth and \ion{H}{i} fractions inferred from the QSO spectra and LAEs observations can be simultaneously achieved solely by high-$z$ galaxies if the escape fraction of ionizing photons from the galaxies ($f_{\rm esc}$) is as high as $\sim 0.2$. 
However the typical escape fraction during the EoR is highly uncertain. 

At intermediate redshift $z\sim3$, Lyman continuum (LyC) photons from galaxies have directly been detected, and the escape fractions have been evaluated as $f_{\rm esc} \sim 0.01- 0.1$ \citep{Steidel01,Iwata09,2015ApJ...810..107M,2016arXiv160508782M,2016arXiv161107038N}, although some of the LyC emitters may be low-$z$ contaminants \citep{2015ApJ...804...17S}. 
In contrast to galaxies at intermediate redshifts, no ionizing photons from
galaxies at the EoR has been detected due to strong attenuation by the
intergalactic medium (IGM). However, \cite{Dunlop13} have claimed that
spectral energy distributions (SEDs) of high-$z$ galaxies imply the
escape fraction of $f_{\rm esc} \approx 0.1-0.2$ with some assumptions. 
The escape fraction also has been assessed by numerical simulations \citep[e.g.][]{Gnedin08, Razoumov10, Yajima11, Paardekooper13, Wise14,2014ApJ...788..121K}. 
Although most of the simulations show a similar trend, i.e. the escape fraction decreases with the mass of the galaxy, there is variance with $\sim$ 1-2 orders of magnitude among the simulation results. 
Thus, there is as yet no consensus on the value of the escape fraction, and this fact leads to the significant uncertainty of the contribution of high-$z$ galaxies to cosmic reionization. 

Active galactic nuclei~(AGNs) are also expected to have been another type of ionizing sources. 
Since photons with energies above $54.4 \rm eV$ are required for doubly ionizing helium, AGNs are the most promising ionizing sources at the epoch when helium in the IGM is fully ionized, i.e.~$z\sim2.7$ \citep{2011MNRAS.410.1096B, 2014arXiv1405.7405W}. 
However, it has been traditionally believed that the contribution of AGNs to reionization is less important because of a rapid decrease in their abundance at $z>3$ \citep{2012ApJ...755..169M, 2014ApJ...786..104U}.  
Very recently \cite{2015A&A...578A..83G} have found faint AGNs at $z\sim
4$-$6$. This suggests that the contribution of AGNs to reionization could
increase by orders of magnitude. 
Although it is still under debate, the contribution of AGNs might overcome that of galaxies if this claim was valid beyond $z\approx6$.  
Unfortunately, while there is currently no direct observation for faint AGNs at $z>6$, the contribution of AGNs could be imprinted on \ion{H}{i} 21cm-line
signal~\citep{2010A&A...523A...4B}.
Hence the Square Kilometer Array~(SKA) and its pathfinders will tell us the information on the abundance of AGNs at $z>6$.

\cite{2015arXiv150707678M} (hereafter MH15) calculated reionization history, extrapolating the evolution of AGN abundance reported by \cite{2015A&A...578A..83G} up to $z=12$ and neglecting the contribution of galaxies. 
As a result, they have found that the ``AGN-dominant scenario'' can satisfy the observed \ion{H}{i}/\ion{He}{ii} fractions and Thomson scattering optical depth simultaneously. 
However, as described above, if $f_{\rm esc}$ is high, the additional contribution of galaxies would results in the overproduction of ionized hydrogen and electron at $z>6$. 
\cite{2015MNRAS.451.1875H} also performed calculations including
both of the galaxies and (faint-)AGNs, and found that
their model is consistent with the EoR measurements,  
the Thomson scattering optical depth and helium reionization.
However, they neglected redshifted high energy photons which
effectively ionize the IGM at lower redshifts, in particular, helium reionization.

\cite{2016arXiv160602719M} investigated contribution of AGNs to reionization history by using their semi-analytic reionization model. 
They have pointed out that their high AGN emissivity model can reionize
\ion{H}{i}.
Their model also leads to more rapid evolution of Lyman limit systems
than in observations, while the reionization evolution by the combination
of lower AGN emissivity and galaxies with the escape fraction of $\sim$ 10\% shows
good agreement with the observation results. 
\cite{2016MNRAS.457.4051K} have numerically shown that abundant faint
AGNs or $f_{\rm esc}$ rising with redshift is required for explaining a
constant photoionization rate at $3<z<5$ reported by observations. 

{In those numerical studies, SEDs of AGNs have been assumed to be simple power-law spectra, however AGNs at $z\sim1$ show SEDs with double peaks at energy ranges of UV~($\sim10~\rm eV$) and X-ray~($\sim10~\rm keV$) \citep[e.g.][]{1997ApJ...475..469Z,1997ApJ...477...93L} that can be theoretically explained by the radiation from a system composed of the accretion disc and coronal gas \citep[e.g.][]{2001ApJ...546..966K}. 
It is well known that the shape of SED controls the efficiency of the ionization and heating processes of the IGM. 
Therefore, the SED shape is an important factor in the calculation for the EoR.
In spite of its importance, the SED shape of AGNs in the reionization era ($z>6$) is highly uncertain especially in the X-ray regime where the heating process effectively contributes to the thermal history. 
Therefore it is worth elucidating how much the SED shape impacts both the reionization and thermal histories. 
}

{In this theoretical work, we aim to constrain the contribution of galaxies and AGNs to cosmic reionization, as well as investigate the effect of the SED shape of AGNs on the reionization and thermal histories. }
For the purpose, we firstly construct a model of the redshift evolution of
the AGN emissivity in which the redshift evolution of the AGN LF and SED depending on the the central black hole mass $M_{\rm BH}$ and the Eddington ratio $\lambda_{\rm Edd}\equiv L/L_{\rm Edd}$ where $L_{\rm Edd}$ is the Eddington luminosity,  are considered. 
Then we compute the H/He reionization and thermal histories with different parameter sets of $f_{\rm esc}$ and AGN abundance. 
{To clarify the impact of the SED shape on the reionization and thermal histories, we employ not only the newly developed SED model but also a simple power-law SED model. } 
Finally, we search possible contribution of galaxies and AGNs to reionization from the comparison between the computed reionization histories and observations. 

The structure of this paper is as follows. 
In section 2, we describe the method for computing the hydrogen and helium reionization histories. 
In section 3, we show our numerical results and compare them to observed \ion{H}{i}/\ion{He}{ii} fractions. 
We discuss the implication of our results and comment on the model
dependence in section 4.
Finally the summary will be given in section 5. 
Throughout this paper, we assume $\rm \Lambda CDM$ cosmology with ($\Omega_{\rm m}, \Omega_{\Lambda},\Omega_{\rm b}, H_0$)=(0.308, 0.692, 0.02237, 67.80km $\rm s^{-1}$ $\rm Mpc^{-1}$)\citep{2015arXiv150201589P}.

\section{Methodology}\label{sec: Methods}
The evolution of the ionization fraction~$f_i$~(for species $i={\rm \ion{H}{i},~\ion{He}{ii},~\ion{He}{iii}}$) can be written as \citep[e.g.][]{1999ApJ...514..648M}
\begin{eqnarray}
	\frac{df_{\rm \ion{H}{ii}}}{dt}&=&\frac{1}{n_{\rm H}(1+z)^3}\frac{dN_{\gamma,\rm \ion{H}{i}}}{dt}-\alpha_{\rm B,\ion{H}{ii}} C
	n_{\rm e}(1+z)^3 f_{\rm \ion{H}{ii}},  	
	\label{eq:ion_1}\\
	\frac{df_{\rm \ion{He}{ii}}}{dt}&=&\frac{1}{n_{\rm He}(1+z)^3}\left(\frac{dN_{\gamma,\rm \ion{He}{i}}}{dt}-\frac{dN_{\gamma,\rm \ion{He}{ii}}}{dt}\right) \nonumber \\
	&-&C n_{\rm e}(1+z)^3 (\alpha_{\rm B,\ion{He}{ii}}  f_{\rm \ion{He}{ii}}-\alpha_{\rm B,\ion{He}{iii}} f_{\rm \ion{He}{iii}})  
	\label{eq:ion_2}\\
	\frac{df_{\rm \ion{He}{iii}}}{dt}&=&\frac{1}{n_{\rm He}(1+z)^3}\frac{dN_{\gamma,\rm \ion{He}{ii}}}{dt}\nonumber \\
 	&-&\alpha_{\rm B,\ion{He}{iii}} C n_{\rm e}(1+z)^3 f_{\rm \ion{He}{iii}},
 	\label{eq:ion_3}
\end{eqnarray}
where $\alpha_{{\rm B},i}$ is the case B recombination coefficient for each species, $n_{\rm H}$, $n_{\rm He}$, and $n_{\rm e}$ are the comoving number densities for hydrogen, helium, and electron, respectively. 
Here $C$ is the clumping factor of the IGM for which we use $C=3$ as shown by 
hydrodynamic simulations taking into account photo-heating of the IGM \citep{Pawlik09,Raicevic2011}. 
{In Eqs. (\ref{eq:ion_1})-(\ref{eq:ion_3}), $dN_{\gamma, i}/dt$ is the photo-ionization rate per unit volume for $i$-th species. 
In this paper, we consider two types of ionizing sources; using the photo-ionization rate per unit volume by star forming galaxies $dN_{*, i}/dt$, and that by AGNs $dN_{{\rm AGN}, i}/dt$, the total photo-ionization rate per unit volume is expressed as $dN_{\gamma, i}/dt = dN_{*, i}/dt+dN_{{\rm AGN}, i}/dt$.}
We will describe how we consider each contribution in the following subsections.

Ionizing photons emitted by stars and AGNs also heat up the IGM. 
{We should note that previous studies (e.g., MH15) omitted the thermal evolution and assumed a fixed temperature. 
However it is expected that the thermal history is sensitive to the shape of SED and AGN abundance. 
Therefore we consistently calculate the thermal and ionization evolution of the IGM.}
The evolution of the IGM temperature $T_{\rm e}$ follows 
\begin{equation}
	\frac{dT_{\rm e}}{dt}=\frac{2}{3}\frac{\mu}{k_{\rm B}n_{\rm B}} \left(\sum_{i}
	\epsilon_{{\rm heat},i} - \epsilon_{\rm cool} \right)  
	-2H T_{\rm e}-{\mu}T_{\rm e} \frac{d\mu^{-1}}{dt} ,
	\label{eq:thermal}
\end{equation}
where $H$ is the Hubble parameter, $k_{\rm B}$ is the Boltzmann
constant, $\mu$ is  the mean molecular weight, $n_{\rm B}$ is the {proper} baryon number density, and $\epsilon_{\rm cool}$ is the IGM cooling rate. 
We take into account the IGM cooling processes summarized in the appendix of~\cite{1994MNRAS.269..563F}.
In Eq.~(\ref{eq:thermal}), $\epsilon_{{\rm heat},i}$ describes the heating rate for each species $i$.
We also decompose the heating rate to two components: from star forming
galaxies, $\epsilon_{*,i}$, and from AGNs, $\epsilon_{{\rm AGN},i}$, as $N_\gamma$ in the case of calculations of the ionization fraction.

When we solve Eqs.~(\ref{eq:ion_1})-(\ref{eq:thermal}), we employ some artificial treatments to bypass some numerical inconveniences. 
As described later, we assume that all ionizing photons from star-forming galaxies are consumed to ionize the local IGM. 
Due to this treatment, $f_{\rm \ion{H}{ii}}$ can exceed 1.0 and $f_{\rm \ion{H}{i}}$ essentially becomes zero. 
Unfortunately, this unrealistic behavior results in numerical oscillations of $f_{\rm \ion{H}{i}}$ and $f_{\rm \ion{H}{ii}}$. 
In order to avoid the unrealistic behavior we set the lower limit of
neutral hydrogen fraction as $f_{{\rm \ion{H}{ii}}, \rm l}=10^{-4}$.
However
this artificial limit leads to the continual heating of the IGM. 
Thus we also set the upper limit of the temperature as $T_{{\rm e}, \rm u}=20,000$K. 

\subsection{Ionizing photons from stars}
In most of previous studies on reionization, star forming galaxies have been considered as main sources of ionizing photons. 
In a star forming galaxy, huge amount of photons are produced from stars, and some fraction of these photons can escape from the galaxy and ionize the surrounding IGM. 
Thus, to evaluate the photo-ionization rate per unit volume by star forming galaxies ($dN_{\ast,i}/dt$), we need to know the value of the escape fraction $f_{\rm esc}$.
However $f_{\rm esc}$ is highly uncertain as described in \S \ref{intro}. 
Therefore, we take $f_{\rm esc}$ as a free parameter in our model. 
Since $f_{\rm esc}$ controls the contribution of high-$z$ galaxies to the cosmic reionization process,
we obtain the constraint on $f_{\rm esc}$ from the comparison with observational data of the EoR. 
For simplicity, we assume that $f_{\rm esc}$ does not depend on the halo mass and redshift. 
Using $f_{\rm esc}$, we assume that the proper number density of photons emitted per unit time from star forming galaxies at a redshift $z$ is proportional to the star formation rate density~(SFRD),
\begin{eqnarray}
	\dot n_{*\nu} = (1+z)^3f_{\rm esc} \dot{\rho}_{*} \gamma_\nu,  
	\label{eq:photon_*}
\end{eqnarray} 
where $\dot{\rho}_*$ is the comoving SFRD for which we adopt the fitting formula by \cite{2014ARA&A..52..415M},
\begin{eqnarray}
	\dot{\rho}_{*}(z)=0.015\frac{(1+z)^{2.7}}{1+[(1+z)/2.9]^{5.6}}\;[{\rm M}_\odot\;{\rm yr^{-1}}{\rm Mpc}^{-3}]. 
\end{eqnarray}
In Eq.~(\ref{eq:photon_*}),
$\gamma_{\nu}$
is the intrinsic number of photons produced by stars per
solar mass at a frequency $\nu$.
For $\gamma_{\nu}$, we use the correspondent formula in~\cite{2005MNRAS.361..577C} that is derived by a population synthesis calculation, 
\begin{eqnarray}
\int^{\nu_2} _{\nu_1} d\nu~ \gamma_\nu =
   \begin{cases}
    5.43 \times 10^{60}/{\rm M_\odot}, & (\nu_1, \nu_2) =(\nu_{\rm \ion{H}{i}},
    \nu_{\rm \ion{He}{i}}), \\
        2.61 \times 10^{60}/{\rm M_\odot}, &(\nu_1, \nu_2) =(\nu_{\rm \ion{He}{i}},
    \nu_{\rm \ion{He}{ii}}), \\
        0.01 \times 10^{60}/{\rm M_\odot}, & (\nu_1, \nu_2) =(\nu_{\rm \ion{He}{ii}},
    \nu_{\rm max,*}).
  \end{cases}
\label{eq:photonnum}
\end{eqnarray}
where $\nu_{\rm max,*}$ is the upper limit frequency of photons from galaxies in our calculation, and is set to $h_{\rm p}\nu_{\rm max,*}=100~{\rm eV}$. Here $h_{\rm p}$ is the Planck constant. 

As mentioned above, the ionizing photons from star forming galaxies are in the range of 13.6 eV to 100 eV. 
Since these photons have short mean free paths, they are consumed to ionize a local region of the IGM. 
This means that all photons from star forming galaxies can be assumed to contribute to ionize the local IGM. 
{We note that this assumption is not reasnable for \ion{H}{i} ionizing photons at lower redshift $z<5$ because the mean free path is order of 100 proper Mpc at $z=3$, shown in Fig.10 of \cite{2014MNRAS.445.1745W}. }
{Following previous studies \cite[e.g.][]{2015MNRAS.451.1875H}, we work with this assumption, but also employ an additional modification that ensures the conservation of the number of ionizing photons. 
With this assumption, a photon is inevitably absorbed by either an \ion{H}{i} atom, a \ion{He}{i} atom, or a \ion{He}{ii} atom. 
Therefore, to calculate photo-ionization rates, it is necessary to evaluate how much extent of photons from star forming galaxies is consumed for the ionization of the $i$-th species otherwise double-counting of ionizing photons occurs. 
The relative number of photons consumed for the ionization of the $i$-th species should be proportional to the relative opacity of the corresponding species \citep[e.g.][]{2006PASJ...58..445S,Umemuraetal2012}.} 
Taking into account the frequency dependence of the ionization cross sections for hydrogen and helium, we can write the photo-ionization rate per unit volume for the $i$-th species as
\begin{eqnarray}
	\frac{dN_{*,i}}{dt} = 
	 \int_{\nu_i}^{\infty}d\nu \frac{ n_{i}\sigma_i \dot{n}_{*\nu}}
	 {n_{\rm \ion{H}{i}} \sigma_{\rm \ion{H}{i}}+n_{\rm \ion{He}{i}} \sigma_{\rm \ion{He}{i}}+n_{\rm \ion{He}{ii}} \sigma_{\rm \ion{He}{ii}}},
	 \label{eq:N_p}
\end{eqnarray}
where $n_i$ and $\sigma_i$ are the number density and cross section for the $i$-th species, respectively.  
The Lyman limit frequency of the $i$-th species is expressed as $\nu_i$, i.e. $h_{\rm p}\nu_{\rm \ion{H}{i}}=13.6$ eV, $h_{\rm p}\nu_{\rm \ion{He}{i}}=24.5$ eV, and $h_{\rm p}\nu_{\rm \ion{He}{ii}}=54.4$ eV.

Next, we consider the heating rate due to radiation from star forming galaxies $\epsilon_{*,i}$.
When a photon with a frequency $\nu$ ionizes a particle of the $i$-th species, the energy $h_{\rm P} (\nu-\nu_i)$ contributes to heat the IGM gas. 
Thus, the photo-heating rate per unit volume $\epsilon_{*,i}$ for the $i$-th species is given by
\begin{eqnarray}
	\epsilon_{*,i}&=& \int_{\nu_i}^{\infty}d\nu \frac{n_i\sigma_i
	 h_{\rm P}(\nu-\nu_{i})\dot{n}_{*\nu}}{n_{\rm \ion{H}{i}} \sigma_{\rm \ion{H}{i}}+n_{\rm \ion{He}{i}}
 	\sigma_{\rm \ion{He}{i}}+n_{\rm \ion{He}{ii}} \sigma_{\rm \ion{He}{ii}}}.
	\label{eq:heatrate}
\end{eqnarray}

As shown by Eq.~(\ref{eq:photonnum}), we only know the integrated $\gamma_\nu$. 
Therefore, we perform the integration in Eqs.~(\ref{eq:N_p}) and (\ref{eq:heatrate}) by using the cross sections averaged over each frequency range. 

\subsection{Ionizing photons from AGNs}
Photons from AGNs are also important in the reionization process of the IGM in the early Universe. 
X-ray photons have much larger mean free paths than UV photons and can ionize and heat up the IGM at cosmological distances, and hence an assumption of ``local ionization'' adopted for star forming galaxies is invalid. 
Therefore, to evaluate the ionization and heating rates at a redshift $z$, we need to take into account X-ray photons coming from AGNs at redshifts higher than $z$. 
Then, the photo-ionization rates and photo-heating rates by AGNs are respectively given by \cite[e.g. ][]{1999ApJ...514..648M}, 
\begin{eqnarray}
	\frac{dN_{{\rm AGN},i}}{dt}(z)=\int_{\nu_{i}}^{\nu_{\rm max,AGN}}
	\frac{d\nu}{h_{\rm P}\nu} \sigma_in_i(1+z)^3 F(z,\nu),
\end{eqnarray}
and 
\begin{eqnarray}
	\epsilon_{{\rm AGN},i}(z)=\int_{\nu_{i}}^{\nu_{\rm max,AGN}}\frac{d\nu}{h_{\rm p}\nu}\sigma_{i}(\nu)
	n_{i} h_{\rm p} (\nu-\nu_{i}) (1+z)^3F(z,\nu).
\end{eqnarray}
where $F(z,\nu)$ represents the specific energy flux from AGNs at redshifts higher than $z$ at  a frequency $\nu$.  
The upper bound of the integration, $\nu_{\rm max,AGN}$, is set to be $2.4\times 10^{20}$ Hz which corresponds to the energy of $10^3$ keV. 
The specific energy flux at a redshift $z$ is calculated as
\begin{eqnarray}
	F(z,\nu)=\int_{z} dz' \varepsilon_{\rm c}(z',\nu')  
	\frac{c (1+z)^2}{H}
	\exp(-\tau_{\nu}(z,z',\nu')),
	\label{F}
\end{eqnarray}
where $c$ is the speed of light and $\nu'=\nu(1+z')/(1+z)$. 
Here, $\varepsilon_{\rm c}(z,\nu)$ is the comoving emissivity, i.e., the specific radiation energy density per unit time coming from AGNs at a redshift $z$ and a frequency $\nu$.
As mentioned in \S~\ref{intro}, previous studies focusing on the contribution of AGNs to reionization \citep[e.g.][]{2015arXiv150707678M,2015MNRAS.451.1875H,2016arXiv160602719M,2016MNRAS.457.4051K} have assumed  power-law spectra up to hard X-ray regime ($\gtrsim 10~\rm keV$) and the evolution of $\varepsilon_{\rm c}$ itself for simplicity. 
However, the shape of SED and the comoving emissivity are highly expected to depend on the evolution of AGNs. 
Therefore, we take into account the evolution of AGN LF and the SED depending on
$M_{\rm BH}$ and $\lambda_{\rm Edd}$ to elucidate the AGN contribution
at the EoR in more detail. 

Using the SED~$L_{\nu}$ and UVLF~$\Phi(z, M_{\rm UV})$ of AGNs, where $M_{\rm UV}$ is the absolute magnitude at the wavelength of 1450~$\AA$, we can write $\varepsilon_{\rm c} (z,\nu)$  
as
\begin{eqnarray}
	\varepsilon_{\rm c}(z,\nu)
 	=\int_{M_{\rm UV}^{\rm min}}^{M_{\rm UV}^{\rm max}}
	 L_{\nu}(M_{\rm UV})\Phi(z, M_{\rm UV})
 	d{M_{\rm UV}}. 
	\label{eq:emis}
\end{eqnarray}
The upper bound of the integration is fixed to be $M_{\rm UV}^{\rm max}=-34$, while the lower bound is set to $M_{\rm UV}^{\rm min}=-18$ as a fiducial value.
The SED ${L_{\nu}}$ is a function of $M_{\rm UV}$. 
We will describe how the $M_{\rm BH}$ and $\lambda_{\rm Edd}$ are related to $M_{\rm UV}$ for performing the integration with respect to $M_{\rm UV}$ (\S~\ref{SEDmodel}), and how the evolution of the UVLF $\Phi$ [$\rm Mpc^{-3} mag^{-1}$] is modeled (\S~\ref{LFmodel}). 
Hereafter, ``UV'' denotes the wavelength of 1450~$\AA$, unless otherwise noted. 

The optical depth for a photon with an observed frequency $\nu$ traveling from $z'$ to $z$, $\tau_{\nu}(z,z',\nu)$, is given by
 \begin{eqnarray}
	\tau_{\nu}(z, z',\nu)&=&\int_{z}^{z'}d\hat{z}
	\frac{c (1+\hat{z})^2}{H}\bigg(  f_{\rm \ion{H}{i}}n_{\rm H}\sigma_{\rm \ion{H}{i}}(\hat{\nu})\nonumber\\
	&+&f_{\rm \ion{He}{i}}n_{\rm He}\sigma_{\rm \ion{He}{i}}(\hat{\nu})+f_{\rm \ion{He}{ii}}n_{\rm He}\sigma_{\rm \ion{He}{ii}}(	
	\hat{\nu}) \bigg)
	\label{eq:tau}
\end{eqnarray}
where $\hat{\nu}=\nu(1+\hat{z})/(1+z)$. 
We assume that the \ion{H}{i} distribution in the Universe is uniform to obtain Eq.~(\ref{eq:tau}), since cosmological simulations usually show that the IGM tends to be uniform with increasing redshift~\citep[e.g.][]{Iliev06}. 
But we should note that the inhomogeneity of the IGM grows during the EoR and might affect the process of the EoR \citep{2016MNRAS.457.4051K}.
Unfortunately, the distribution of Lyman limits systems at the EoR, which contributes to the inhomogeneity of the IGM, is still too unknown to appropriately consider in this study \citep[but see also][]{Inoue 2014}. 
We will discuss the impact of the IGM inhomogeneity on our results in \S~\ref{discussion}. 

\subsubsection{AGN luminosity function}\label{LFmodel}
In this paper, we construct an AGN UVLF model consistent with recent observations. 
According to~\cite{2009MNRAS.399.1755C}, we express the UVLF in the shape of double power-law as
\begin{eqnarray}
	\Phi(z,M_{\rm UV})=\frac{\Phi_{*}\exp(-z/z_{\rm
	 AGN})}{10^{0.4(\alpha+1)(M_{\rm
	 UV}-M_{*})}+10^{0.4(\beta+1)(M_{\rm UV}-M_{*})}}.
	 \label{eq:phizm}
\end{eqnarray}
Here the parameters $\Phi_{*}$, $M_{*}$, $\alpha$ and $\beta$ respectively control the amplitude, characteristic $M_{\rm UV}$ where the slope switches, faint-end slope, and bright-end slope of the UVLF.  
These parameters are set to reproduce observed UVLFs at $z<4$~\citep{2009MNRAS.399.1755C,2015A&A...578A..83G}. {We use $\alpha=-1.5$ at $z=2$ \citep{2009MNRAS.399.1755C}.} In Eq.~(\ref{eq:phizm}), the decrease in AGN number density at high-$z$ is controlled by $z_{\rm AGN}$ for which we assume $z_{\rm AGN}=6$. 

\begin{figure*}
	\centering
	\includegraphics[width=5.5cm]{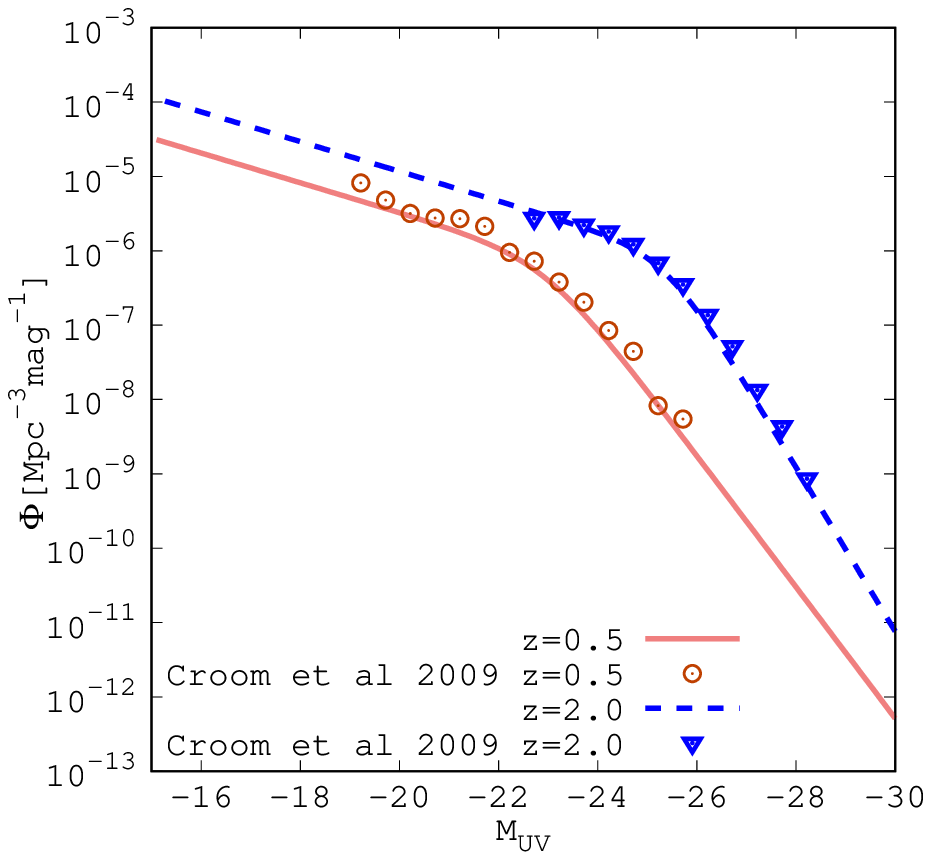}
	\includegraphics[width=5.5cm]{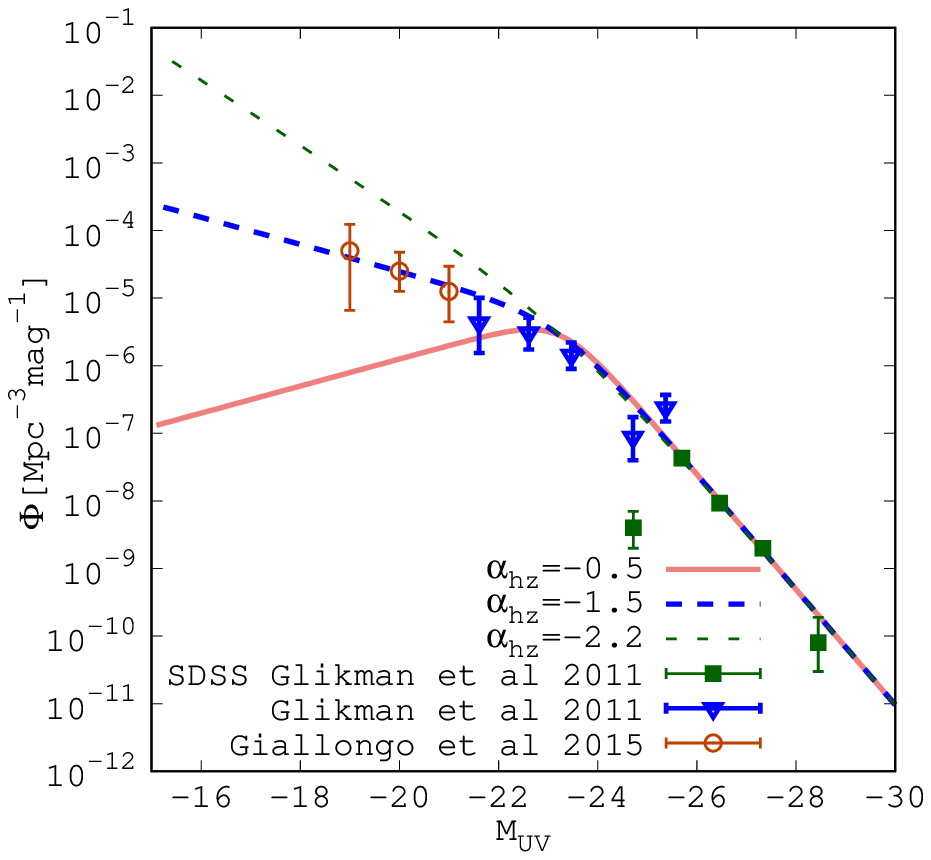}
	\includegraphics[width=5.5cm]{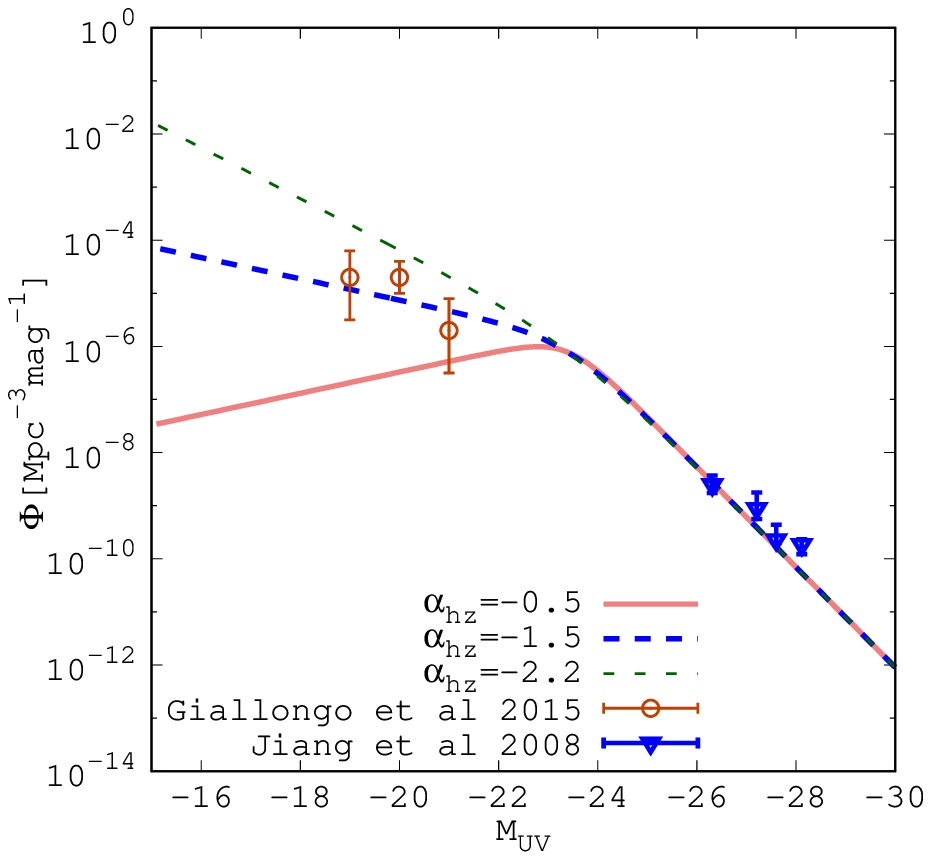}
	\caption{AGN UVLFs at $z<2$ (left), $z=4.25$ (center), and 
	$z=5.75$ (right).  Observational data are taken from 
	\protect\cite{2008AJ....135.1057J,2009MNRAS.399.1755C,2011ApJ...728L..26G,2015A&A...578A..83G}. 
	{In the left panel, the red solid line and the blue dashed line respectively represent 
	the models of LFs at $z=0.5$ and $z=2.0$ with $\alpha_{\rm hz}=-1.5$.} 
	In the central and right panels, modeled LFs with $\alpha_{\rm hz}=0.0$, -1.5, 
	and -2.2 are respectively shown by the red thick solid, blue thick dashed, 
	and green thin dashed lines. }
\label{LFs} 
\end{figure*}

The details of UVLFs at higher redshifts are still under debate, especially in their faint-end parts.
Our aim in this paper is to evaluate these contributions to the EoR process.
Therefore, we introduce a parameter $\alpha_{\rm hz}$ as
the faint-end slope parameter at $z>4.25$.
Then we interpolate $\alpha$ between $z=2$ and $z=4.25$.
For other UVLF parameters, i.e. $\Phi_{*}$, $M_{*}$, and $\beta$ in high
redshifts $z>4.25$ we take the $\log (1+z)$ interpolation using
the values at redshifts $z=4.25$, $z=4.75$ and $z=5.75$ represented in \cite{2015A&A...578A..83G}. We further extrapolate these parameters for $z>5.75$. The UVLF parameters used in this paper are listed in Table~\ref{table1}. We show our AGN UVLF models for different redshifts at Fig.~\ref{LFs}. Our UVLF models are well consistent with observations at each redshift. 
As shown by the right panel of Fig.~\ref{LFs}, the contribution of faint AGNs to reionization increases with decreasing $\alpha_{\rm hz}$ and increasing $M^{\rm min}_{\rm UV}$. 

\begin{table}
  \centering
  \begin{tabular}{|l|c|c|r||r|} \hline
    z & $\alpha$ &$\beta$ & $M_{*}$ & $\log\Phi_{*}$ \\ \hline \hline
    2.00 & -1.5 & -3.76 & -25.4 & -5.86 \\
    4.25 & $\alpha_{\rm hz}$ & -3.13& -23.2 & -4.89 \\
    4.75 & $\alpha_{\rm hz}$ & -3.14& -23.6 & -5.36 \\
    5.75 & $\alpha_{\rm hz}$ & -3.35 & -23.4 & -5.38 \\
    8.00 & $\alpha_{\rm hz}$ & -3.73 & -23.0 & -5.43 \\
    10.0 & $\alpha_{\rm hz}$ & -3.99 & -22.8 & -5.47 \\ \hline
  \end{tabular}
  \caption{Parameters of our AGN UVLF model at $z=2.0$, 4.25, 4.75, 5.75, 8.00, and 10.0. The faint-end slope $\alpha_{\rm hz}$ at $z>4.25$ is a parameter. The $\log \Phi_{*}$ is for $z_{\rm AGN}=6.0$. }
  \label{table1}
\end{table}

\subsubsection{AGNs spectral energy distribution}\label{SEDmodel}
{
In this section, we derive an SED model depending on $M_{\rm BH}$ and
$\lambda_{\rm Edd}$ based on results by \cite{2001ApJ...546..966K} who
have numerically computed the SED of a system composed of the standard accretion disk and coronal gas. 
{Hereafter we call this model as the AC model~(accretion disk + corona). }
We firstly introduce a shape function $\phi(E)$ as
\begin{eqnarray}
	\nu L_\nu  = L_{\rm UV} \phi(E), 
\end{eqnarray}
\begin{eqnarray}	
	\phi(E) = \left\{\bigg(\frac{E}{E_1}\bigg)^{\gamma_a}+\bigg(\frac{E}{E_1}\bigg)^{\gamma_b}\right\}^{-1}+A_c \left\{ \bigg(\frac{E}{E_2}\bigg)^{\gamma_c}+\bigg(\frac{E}{E_2}\bigg)^{\gamma_d}\right\}^{-1},
\label{eq:spectral}
\end{eqnarray}
where $L_{\rm UV}, ~A_c,~\gamma_a,~\gamma_b,~\gamma_c,~\gamma_d,~E_1$,and $E_2$ are the model parameters that are selected for reproducing the SED with $M_{\rm BH}=3\times10^9$ $\rm M_{\odot}$ and $\lambda_{\rm Edd}=0.05$ obtained in \cite{2001ApJ...546..966K}\footnote{To be exact, there is a degeneracy between $\lambda_{\rm Edd}$ and the radiation efficiency $\eta$. We implicitly assume $\eta=0.1$ for the fiducial model.}. 
Carrying out the $\chi^2$ fitting, the parameters are found to be $A_c=0.397$, $\gamma_a =-0.0191$, $\gamma_b=0.685$, $\gamma_c=-0.255$, $\gamma_d=0.211$,  $E_1=9.44\times10^{3}~{\rm keV}$, $E_2=1.22\times10^{-3}~{\rm keV}$, and $L_{\rm UV}=3.4\times 10^{45}~{\rm erg/s}$. 
}

\begin{figure}
\centering
	\includegraphics[width=7cm]{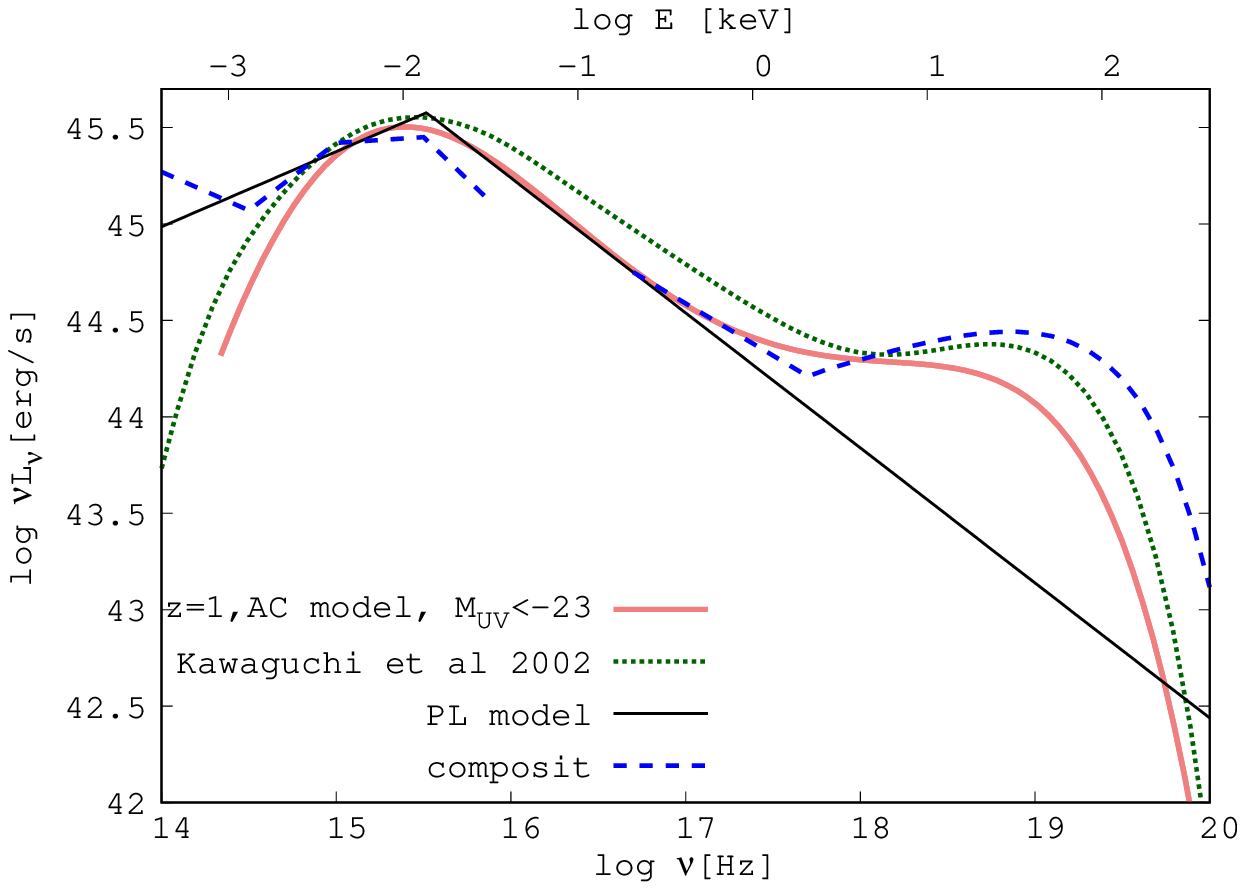}
	\includegraphics[width=7cm]{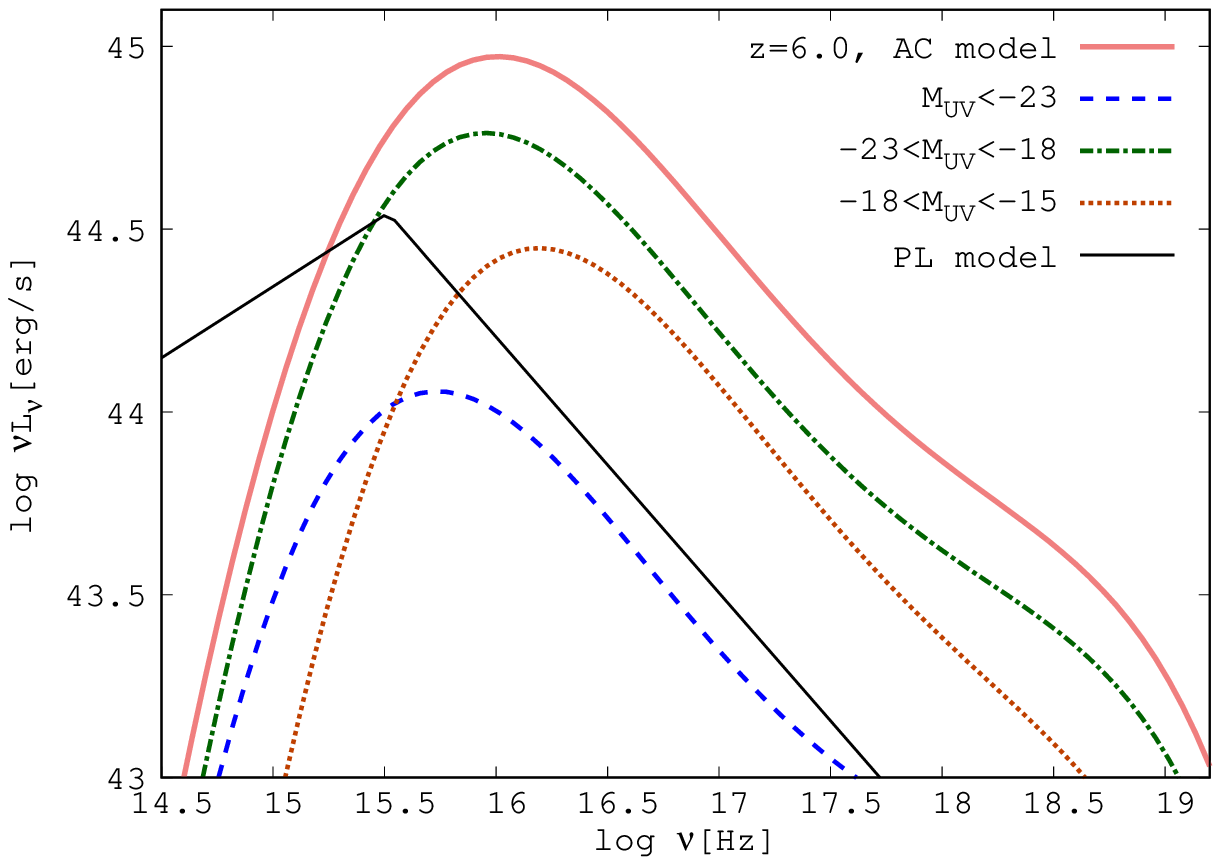}
	\caption{{\it Top}: Weighted average SED for AGNs at $z=1$. The red thick solid and 
	black thin curves respectively represent the AC and PL models. 
	The SED with $M_{\rm BH}=3\times10^9$, 
	$\rm M_{\odot}$ and $\lambda_{\rm Edd}=0.05$ computed by 
	\protect\cite{2001ApJ...546..966K} (green dotted curve), 
	and the observed composite SED \citep{1997ApJ...475..469Z,1997ApJ...477...93L} 
	(blue dashed curve) are also plotted for reference. 
	{\it Bottom}: Weighted average SED for AGNs at $z=6$ with $\alpha_{\rm hz}=-1.5$. 
	The AC model and the PL model are shown by the red thick and black thin solid curves, 
	respectively. {The red solid line is obtained by performing the integrations in Eq. (21) over $-32< M_{\rm UV} < -15$.}
	The blue dashed, green dot-dashed, and orange dotted curves respectively 
	represents the contribution from AGNs with $M_{\rm UV}<-23$, with $-23\le M_{\rm UV} 
	< -18$, and with $-18 \le M_{\rm UV} < -15$. 
	{The sum of these three curves corresponds to the red solid curve.} 
	We note that the contribution from faint 
	AGNs is controlled by $\alpha_{\rm hz}$.}  
\label{sed} 
\end{figure}

We employ theoretically motivated scaling relations shown in~\cite{2001ApJ...546..966K} to take the dependence of the SED on $M_{\rm BH}$ and $\lambda_{\rm Edd}$ into account. According to the standard accretion disk model, the peak frequency of the big blue bump~
(the bump at $E \sim 10~\rm eV$ in Fig.~\ref{sed}) follows the scaling relation, 
\begin{eqnarray}
	\nu_{\rm p}\propto M_{\rm BH}^{-1/4}\lambda_{\rm Edd}^{1/4}. 
	\label{Lumi2}
\end{eqnarray}
In addition, the scaling of the total luminosity is simply given by 
\begin{eqnarray}
	L = \lambda_{\rm Edd}L_{\rm Edd}\propto \lambda_{\rm Edd}M_{\rm BH}. 
	\label{Lumi1}
\end{eqnarray} 
\cite{2001ApJ...546..966K} have indeed shown that these simple relations approximately explain the behavior of their computed SEDs.

{
Owing to the scaling relations, we can uniquely convert $L_\nu(M_{\rm
BH},\lambda_{\rm Edd})$ into $L_{\nu}(M_{\rm UV})$ appearing in Eq.~(\ref{eq:emis}). 
For the redshift evolution of $\lambda_{\rm Edd}$, we adopt $\lambda_{\rm Edd}(z)={\rm min}[0.0091\times (1+z)^{1.9}$,~1.0]~based on observations by \cite{2012ApJ...746..169S}.  
}

{
In order to discuss the impact of the shape of the SED on the reionization and thermal histories, we also perform calculation with a simple power-law SED
{(hereafter the PL model)} proposed by \cite{2015MNRAS.449.4204L}, 
\begin{eqnarray}
L_{\nu} \propto
   \begin{cases}
   \nu^{-0.61}, & h_{\rm p}\nu < 13.6~{\rm eV},\\
   \nu^{-1.70}, & h_{\rm p}\nu \ge13.6~{\rm eV}, 
  \end{cases}
\label{eq:Lusso}
\end{eqnarray}
which is employed in~\cite{2015arXiv150707678M}. 
Also \cite{2016MNRAS.457.4051K} assumed a power-law SED which has different power.
The amplitude of the PL model is determined to match its luminosity at the wavelength of $1450\AA$ to that of the AC model.
{We use Eq.~(\ref{eq:Lusso}) up to X-ray regime ($> 0.5{\rm keV}$) for the PL model as in MH15, but we should mention that broken power-law spectra with the power-law index switching to $\approx -1$ at the X-ray regime are also proposed \cite[e.g.,][]{1993A&A...275....1H}. 
With this type of broken power-law spectra, a larger amount of hard X-ray photons can be produced as similar to the AC model, and thus a broken power-law spectrum should be regarded as an alternative of the AC model \cite[e.g., see Fig.~1 in][]{1997ApJ...477...93L}. 
At present it is still unclear which of these spectra, i.e., the AC model or the broken spectra, is suitable for reproducing the SEDs of high-$z$ AGNs.  
We adopt the AC model for comparison with the simple power-law model used in the previous studies nevertheless, because the AC model and the broken power-law model are expected to yield  similar thermal histories if hard X-ray luminosities are almost identical. }

To compare our SED models with observed SEDs, we calculate the weighted average SED as 
\begin{eqnarray}
	\overline{L_{\nu}}(\nu,z)=\frac{\int \Phi(M_{\rm UV},z) L_{\nu}(M_{\rm UV}) dM_{\rm UV}}{\int 	\Phi(M_{\rm UV},z) dM_{\rm UV}}. 
	\label{aveSED}
\end{eqnarray}
{In the top panel of Fig.~\ref{sed}, we show the SEDs for the AC/PL models at $z$=1, the SED computed by \cite{2001ApJ...546..966K} and observed composite SED \citep{1997ApJ...475..469Z,1997ApJ...477...93L}. 
The figure shows that all SEDs are well consistent with each other at $E\sim10^{-2}~{\rm keV}$, though the composite SED proposed by \citet{1997ApJ...475..469Z} is slightly lower than the others due to less IGM correction \citep{2015MNRAS.449.4204L}.
On the other hand at the X-ray regime ($E>0.5{\rm keV}$), the AC model shows excess coming from the coronal gas compared to the PL model. 
Additionally, in the bottom panel of Fig.~\ref{sed}, we find that the AC model has significantly higher ionizing photon emissivity at $z=6$ due to the dependence of the peak frequency on $M_{\rm BH}$ and
$\lambda_{\rm Edd}$ shown in Eq.~(\ref{Lumi2}). 
}}

\begin{figure}
\centering
	\includegraphics[width=7cm]{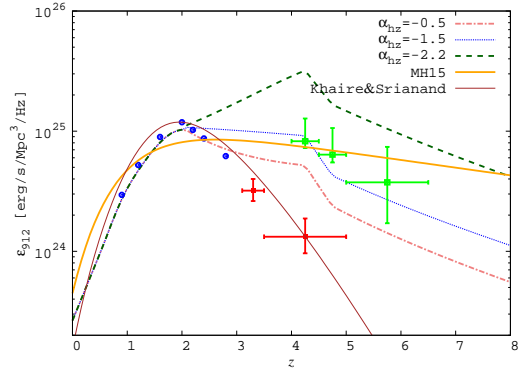}
	\caption{Redshift evolution of AGN emissivity at $912\AA$. 
	The plotted values are converted from the emissivity at $1450\AA$ by using 	
	Eq.~(\ref{eq:Lusso}). 
	The dot-dashed, dotted, and dashed lines respectively represent the emissivity 
	for our UVLF model with $\alpha_{\rm hz}=$-0.5,~-1.5 and -2.2. 
	For comparison, the model in MH15 is shown by the thick solid line and the 
	emissivity obtained by optical measurements in \citet{2015MNRAS.451L..30K} 
	by the thin solid line. 
	We also show AGN ionizing emissivity inferred from other observations: \citet{2012ApJ...755..169M} in red, \citet{2015A&A...578A..83G} in green and \citet{2013A&A...551A..29P} in blue.}
\label{Emis}
\end{figure}

\subsubsection{AGNs emissivity at 13.6~eV}
The AGN hydrogen ionizing emissivity~$\varepsilon_{912}$, namely the emissivity of AGNs at $E=13.6~\rm eV$, is a good indicator to show how much photons emitted from AGNs contribute to cosmic reionization. 
Thus, we firstly compare our model with other models in previous studies by using $\varepsilon_{912}$. 
We plot the redshift evolution of $\varepsilon_{912}$ for different $\alpha_{\rm hz}$ in Fig.~\ref{Emis}. 
The corresponding value in MH15 and \cite{2015MNRAS.451L..30K} are shown by the thick solid and thin solid lines, respectively. 
The luminosity of AGNs over the wide redshift range are obtained through many independent observations.
Using  Eqs.~(\ref{eq:emis}) and (\ref{eq:Lusso}), we also calculate $\varepsilon_{912}$ for different observations. 
\cite{2013A&A...551A..29P} provided g-band luminosity functions in the redshift range, $0.68<z<4$, from SDSS-$\rm I\hspace{-.1em}I\hspace{-.1em}I$ and Multiple Mirror Telescope data. 
{We convert the g-band LFs into i-band LFs by using Eq. (2) in \cite{2009MNRAS.399.1755C} and convert the i-band LFs into UVLFs by using Eq. (B1) in \cite{2015MNRAS.449.4204L}; 
\begin{eqnarray}
	M_{g}(z=2) = M_{i} - 1.25 \log\left(\frac{4670\AA}{7471\AA}\right)\nonumber \\
	M_{i}(z=2) = M_{\rm UV}+1.28,
	\label{convert}
\end{eqnarray}
where $M_{g}$ is the g-band magnitude, $M_{i}$ is the i-band magnitude. {It should be mentioned that different power law indexes are assumed in \cite{2009MNRAS.399.1755C} and \cite{2015MNRAS.449.4204L}. However, the difference in $M_{\rm UV}$ caused by employing the different power law indexes is almost negligible ($\Delta M_{\rm UV}=$0.056). }
We show $\varepsilon_{912}$ inferred from~\cite{2013A&A...551A..29P} by blue circles in Fig.~\ref{Emis}.}
The red points in Fig.~\ref{Emis} show $\varepsilon_{912}$ at $z>3$ inferred from quasar luminosity functions obtained from the Cosmic Evolution Survey (COSMOS) \citep{2012ApJ...755..169M} by which a characteristic decrease of the quasar number density from $z$=3 to 4 is suggested. 
The green squares in Fig.~\ref{Emis} are obtained from the UVLFs of faint AGNs at $4<z<6.5$ found by \cite{2015A&A...578A..83G}. 

As shown in Fig.~\ref{Emis} the AGN emissivity in $z\approx4$ has a large uncertainty; although \cite{2015A&A...578A..83G} used only 22 AGN candidates in the CANDLES GOODS-South field as samples, their result is obviously different from that by \cite{2012ApJ...755..169M}. 
This inconsistency is probably caused by the different sample sets for each study. 
Furthermore, the uncertainty in the AGN emissivity increases at $z>6$ since the number of observed high-$z$ AGNs decreases with increasing redshift. 
Due to the large uncertainty, the measurement of the AGN ionizing emissivity cannot put a strong constraint on the evolution factor~$\alpha_{\rm hz}$. 

\begin{figure}
\centering
	\includegraphics[width=7cm]{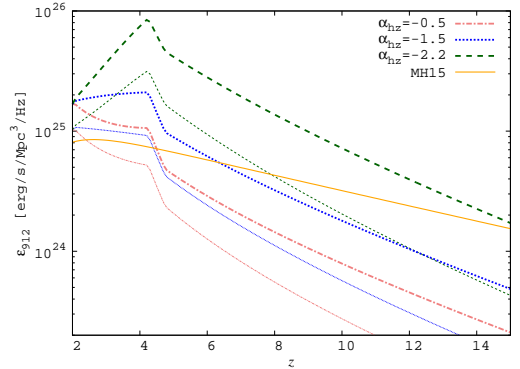}
	\caption{Redshift evolution of AGN emissivity at $912\AA$ for AC model (thick 
	lines) and PL model (thin lines) with $\alpha_{\rm hz}=-0.5$(dot-dashed), 
	-1.5(dotted) and -2.2(dashed). 
	The solid line shows the model in MH15. 
	This figure indicates that $\varepsilon_{912}$ in the extremely AGN abundant model 
	($\alpha_{\rm hz}=-2.2$) are more than an order of magnitude higher than 
	that in MH15. }
\label{EmisSED} 
\end{figure}

Before the comparison with the EoR observation data in next section, we show the AGN emissivity at $z<15$ for different $\alpha_{\rm hz}$ with the AC and PL models.
Fig.~\ref{EmisSED} shows that thick lines corresponding to the emissivity for the AC model are larger than thin lines corresponding to the emissivity for the PL model. 
For reference, we also show the AGN dominant model in MH15 by the dot-dashed line. 
Fig.~\ref{EmisSED} also presents that $\varepsilon_{912}$ in the extremely AGN abundant model ($\alpha_{\rm hz}=-2.2$) are more than an order of magnitude higher than that in MH15. 
This fact suggests that the photo-ionization rates in the extreme models exceed observationally measured values at $2<z<6$  \cite[e.g.,][]{Calverley11, 2013MNRAS.436.1023B} which are consistently reproduced by MH15. 
We discuss this point later in \S~4.3.   

\section{Results}\label{sec:results}

\begin{figure}
\centering
	\includegraphics[width=7cm]{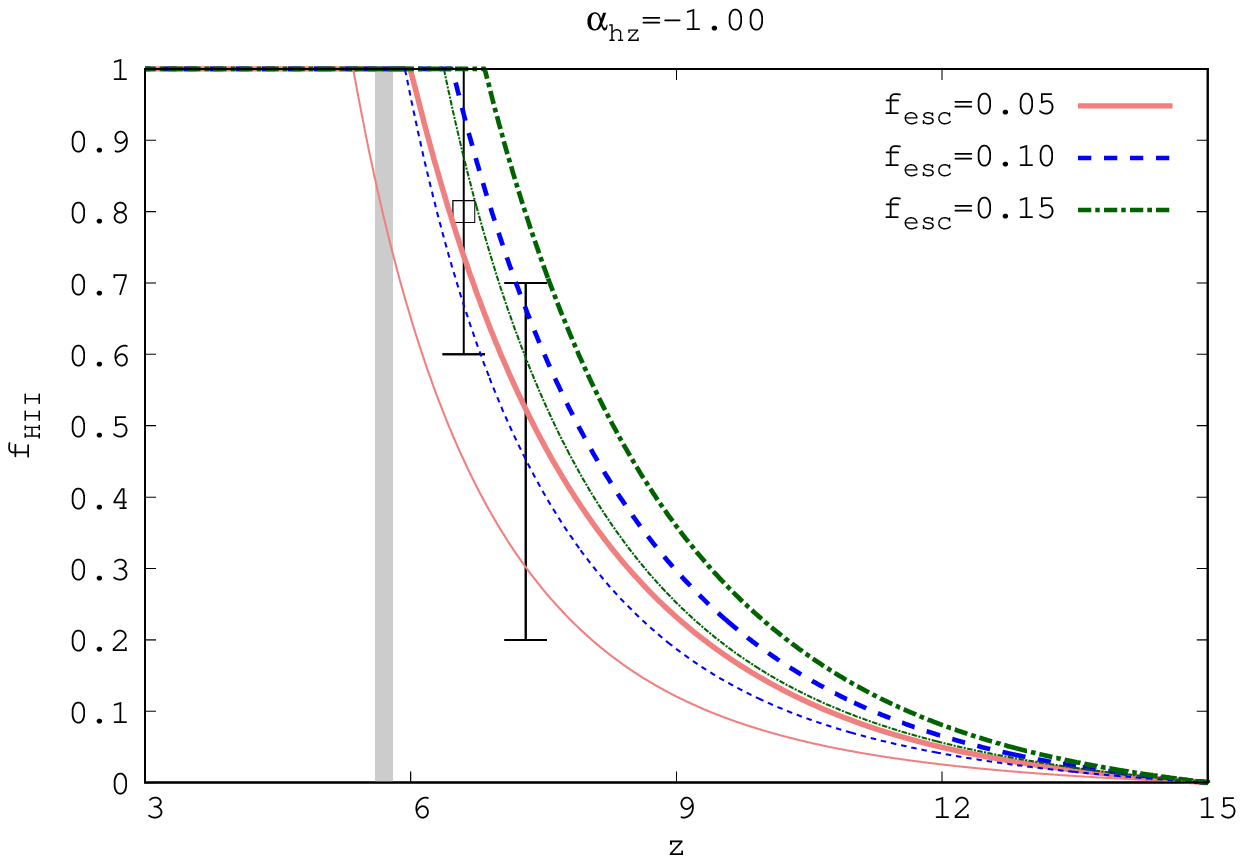}
	\includegraphics[width=7cm]{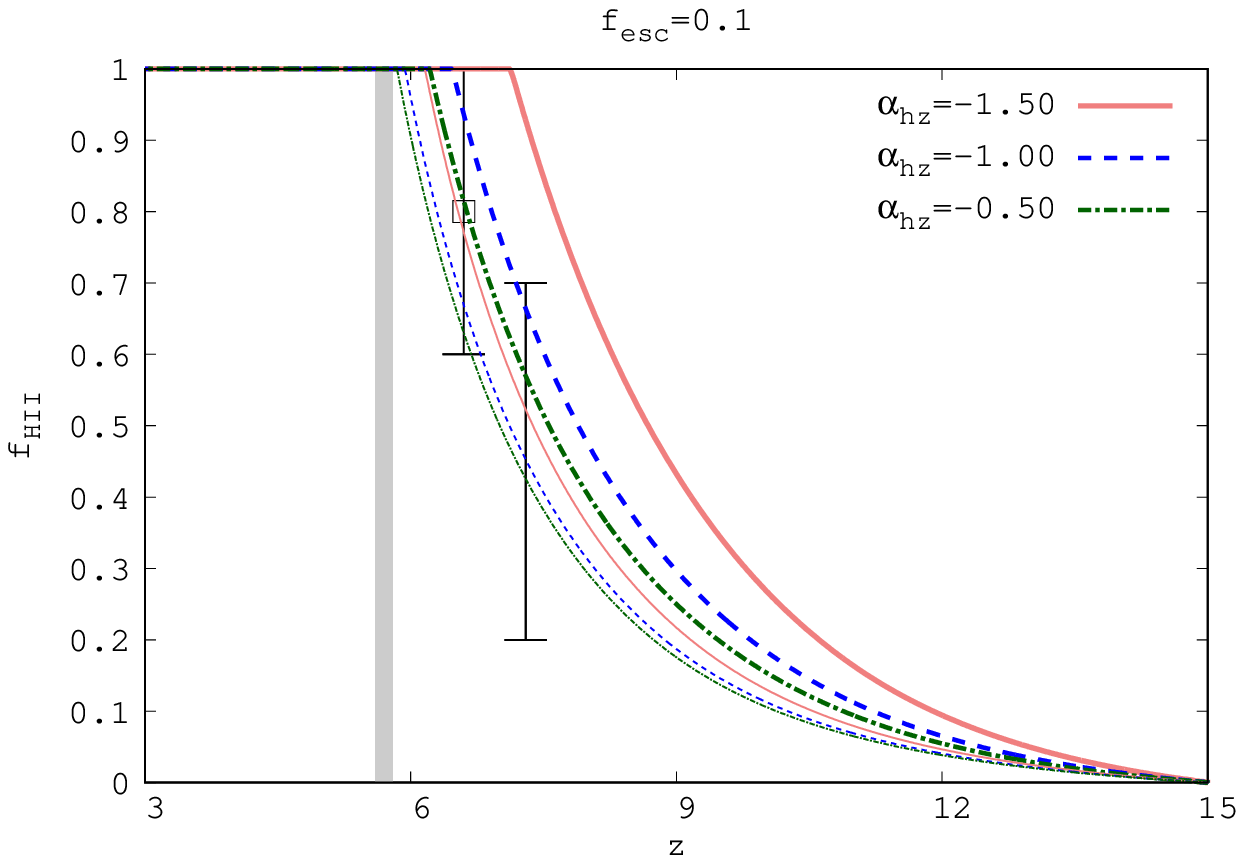}
	\caption{\ion{H}{i} reionization histories computed with various sets of $f_{\rm esc}$ and 
	$\alpha_{\rm hz}$. 
	The thick lines represent \ion{H}{i} reionization histories computed with the AC 
	model, and the thin lines represent those with the PL model. 
	{\it Top panel}: $\alpha_{\rm hz}$ is fixed to be -1.0. The results with 
	$f_{\rm esc}=0.05$, 0.1 and 0.15 are shown by the solid, dashed, and 
	dot-dashed lines, respectively. 
	{\it Bottom panel}: $f_{\rm esc}$ is fixed to be 0.1. The results with 
	$\alpha_{\rm hz}=-1.5$, -1.0 and 0.5 are shown by the solid, dashed, and 
	dot-dashed lines, respectively. 
	In both panels, the gray vertical line corresponds to $z=5.7$ that is 
	the completion 
	redshift of the \ion{H}{i} reionization inferred from observed QSO spectra
	\citep{2006AJ....132..117F}. 
	Two data points with error bars represent the constraints imposed by 
	LAE observations \citep{2010ApJ...723..869O,2014ApJ...797...16K}. 
	The \ion{H}{i} reionization history tends to be completed earlier with increasing 
	$f_{\rm esc}$ and decreasing $\alpha_{\rm hz}$. 
	Besides, for a fixed parameter set of  $f_{\rm esc}$ and $\alpha_{\rm hz}$,  
	\ion{H}{i} reionization with the AC model is always completed earlier than the PL 
	model. } 
\label{IH} 
\end{figure}

\begin{figure}
\centering
	\includegraphics[width=7cm]{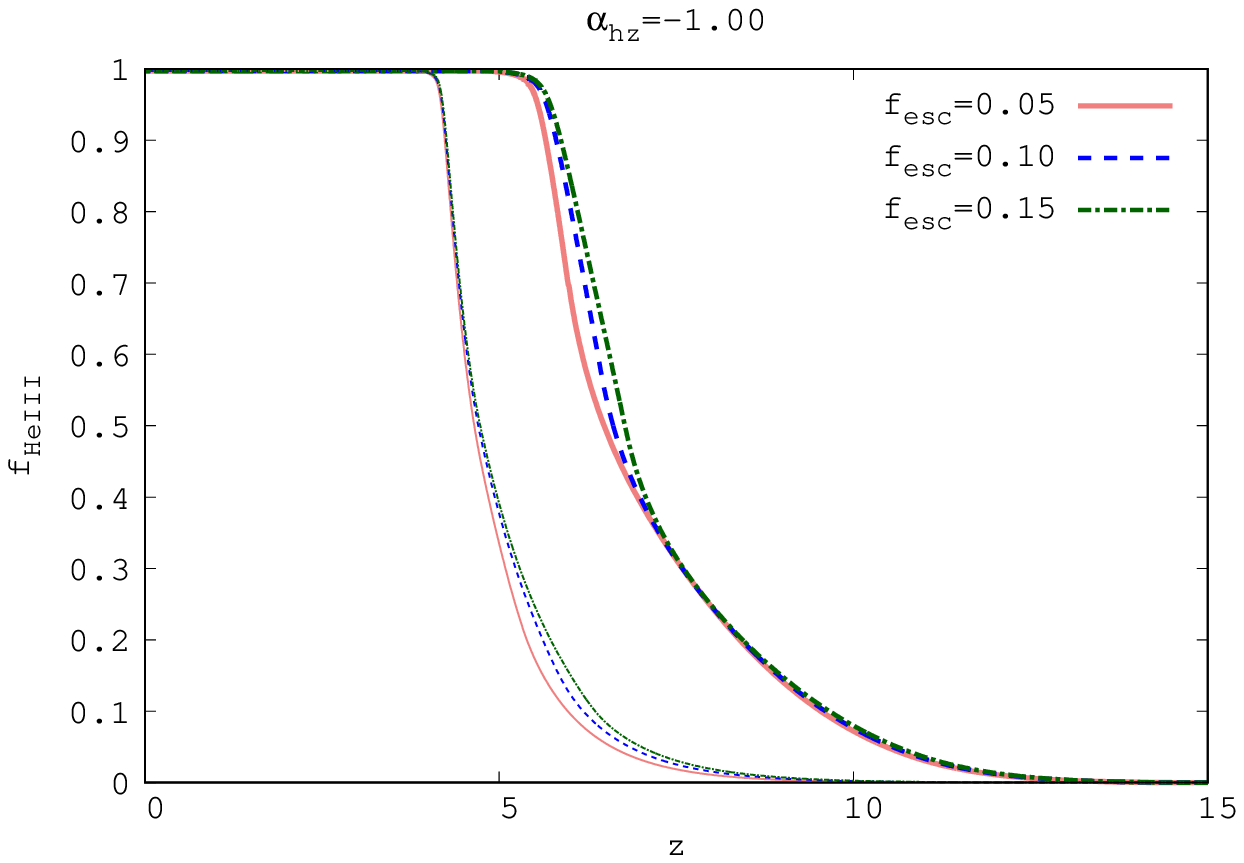}
	\includegraphics[width=7cm]{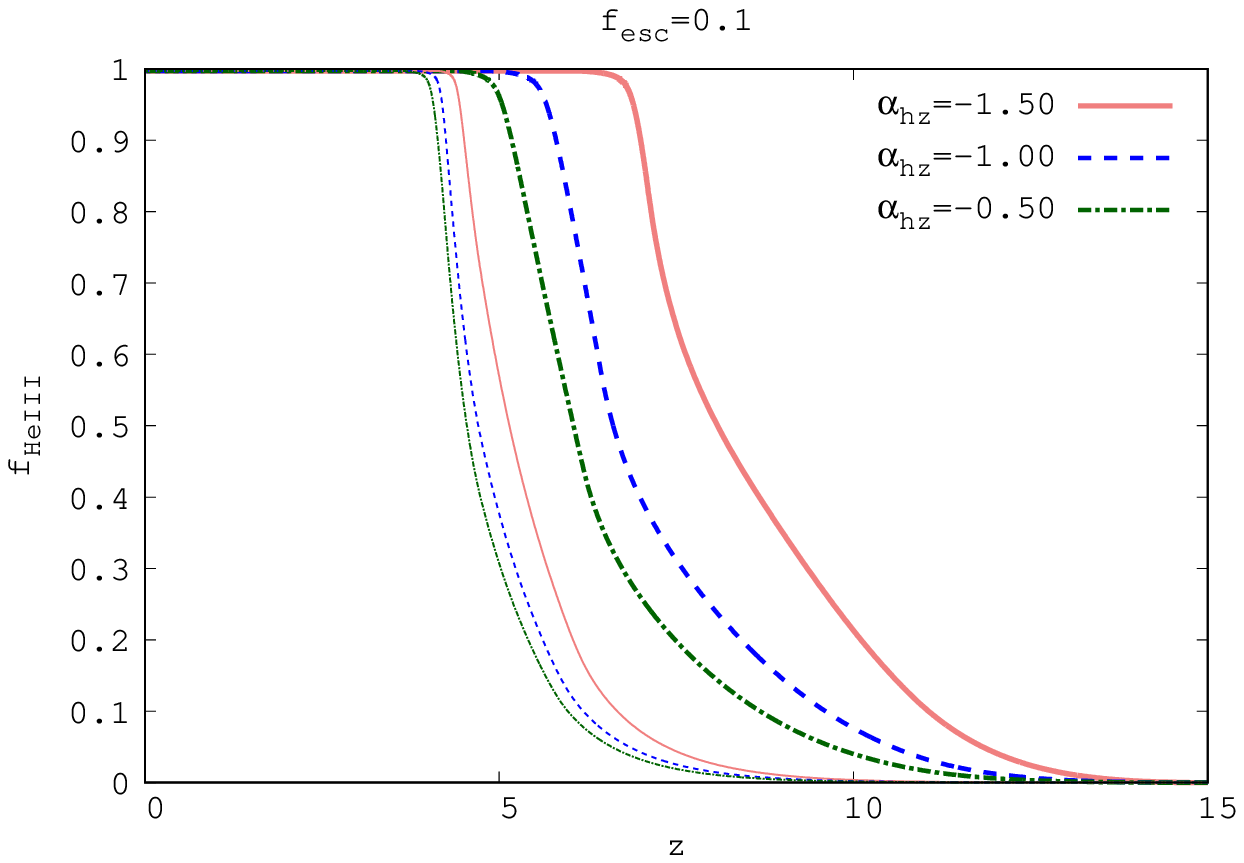}	
	\caption{\ion{He}{ii} reionization histories computed with various sets of 
	$f_{\rm esc}$ and $\alpha_{\rm hz}$. 
	The parameter sets in both panels are the same as those in 
	Fig.~\ref{IH}; 
	the thick lines represent \ion{He}{ii} reionization histories computed with 
	the AC model, and the thin lines represent those with the PL model. 
	{\it Top panel}: $\alpha_{\rm hz}$ is fixed to be -1.0. The results with 
	$f_{\rm esc}=0.05$, 0.1 and 0.15 are shown by the solid, dashed, and 
	dot-dashed lines, respectively. 
	{\it Bottom panel}: $f_{\rm esc}$ is fixed to be 0.1. The results with 
	$\alpha_{\rm hz}=-1.5$, -1.0 and 0.5 are shown by the solid, dashed, and 
	dot-dashed lines, respectively. 
	Similar to the \ion{H}{i} reionization history, the \ion{He}{ii} reionization history 
	tends to be completed earlier with increasing 
	$f_{\rm esc}$ and decreasing $\alpha_{\rm hz}$ although the dependence on 
	$f_{\rm esc}$ is significantly weak. 
	This result straightforwardly indicates that AGNs dominantly contributes to 
	the \ion{He}{ii} reionization. 
	In addition, it turns out that the SED shape of AGNs makes a strong impact on 
	the \ion{He}{ii} reionization history. } 
\label{IHe} 
\end{figure}

\begin{figure}
\centering
	\includegraphics[width=7cm]{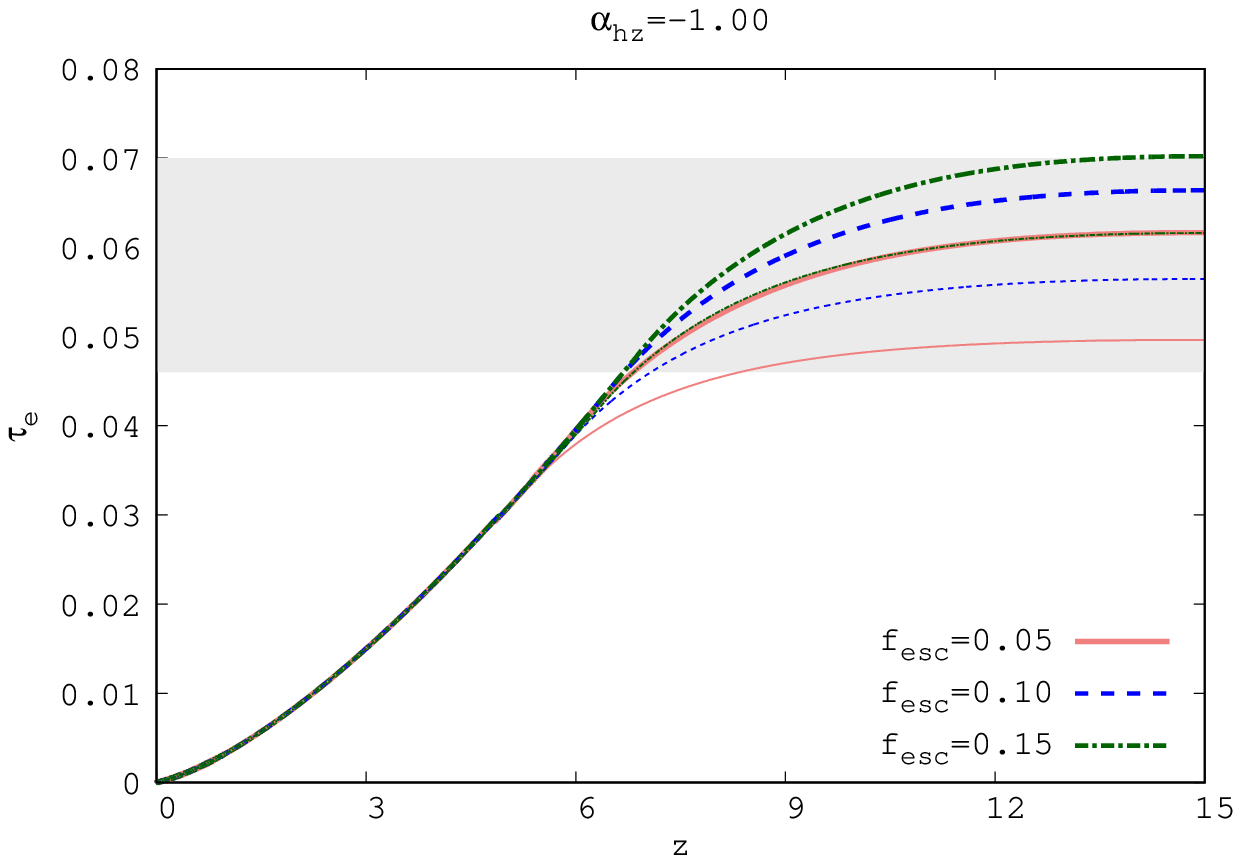}
	\includegraphics[width=7cm]{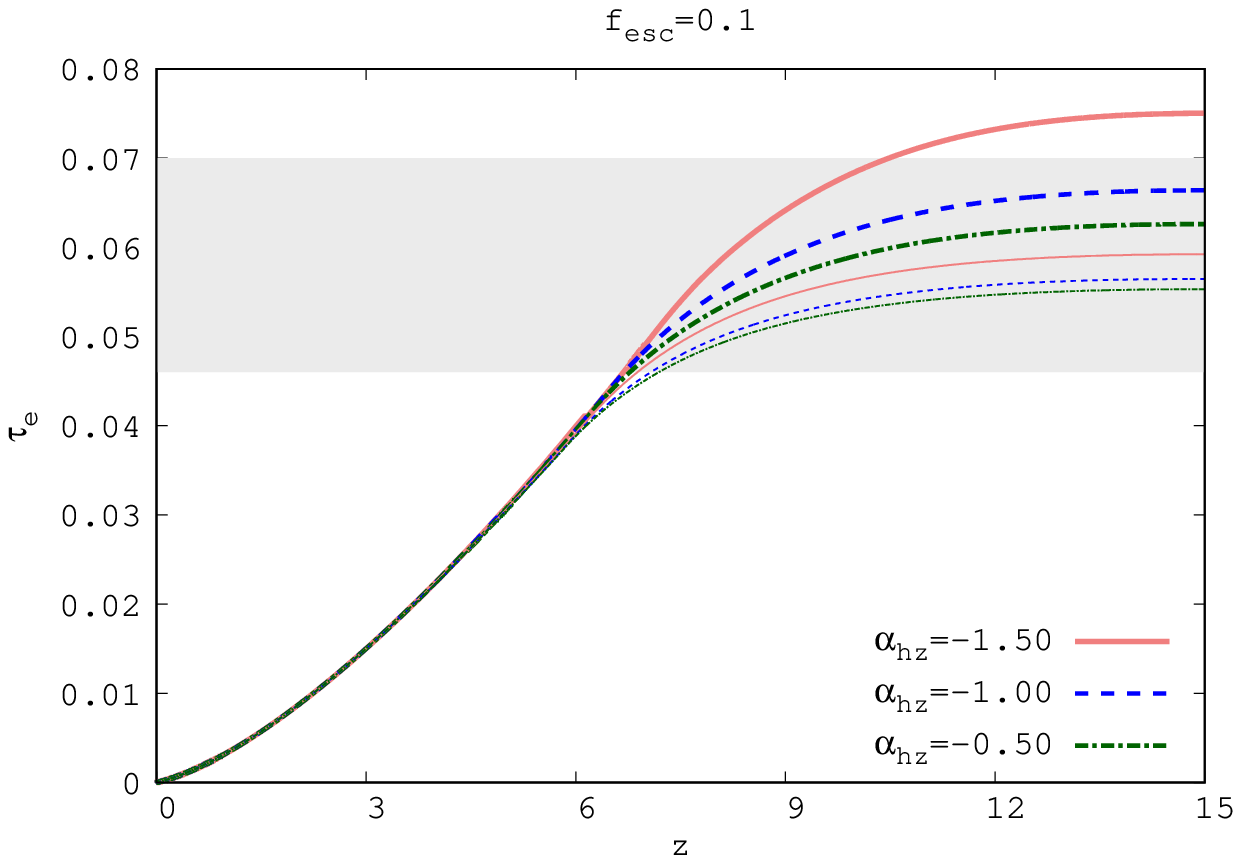}
	\caption{CMB Thomson optical depth with various sets of $f_{\rm esc}$ and 
	$\alpha_{\rm hz}$. 
	The lines and the parameter sets in both panels are the same as those in 
	Fig.~\ref{IH}; 
	the thick lines represent $\tau_{\rm e}$ computed with the 
	AC model, and the thin lines represent those with the PL model. 
	{\it Top panel}: $\alpha_{\rm hz}$ is fixed to be -1.0. The results with 
	$f_{\rm esc}=0.05$, 0.1 and 0.15 are shown by the solid, dashed, and 
	dot-dashed lines, respectively. 
	{\it Bottom panel}: $f_{\rm esc}$ is fixed to be 0.1. 
	The results with $\alpha_{\rm hz}=-1.5$, -1.0 and 0.5 are shown by the solid, 
	dashed, and dot-dashed lines, respectively. 
	In each panel, a constraint imposed by the CMB observation, i.e. 
	$\tau_{\rm e} = 0.058 \pm 0.012$, is shown as a gray region \citep{2016arXiv160503507P}. } 
\label{tau} 
\end{figure}

\begin{figure}
\centering
	\includegraphics[width=8.5cm]{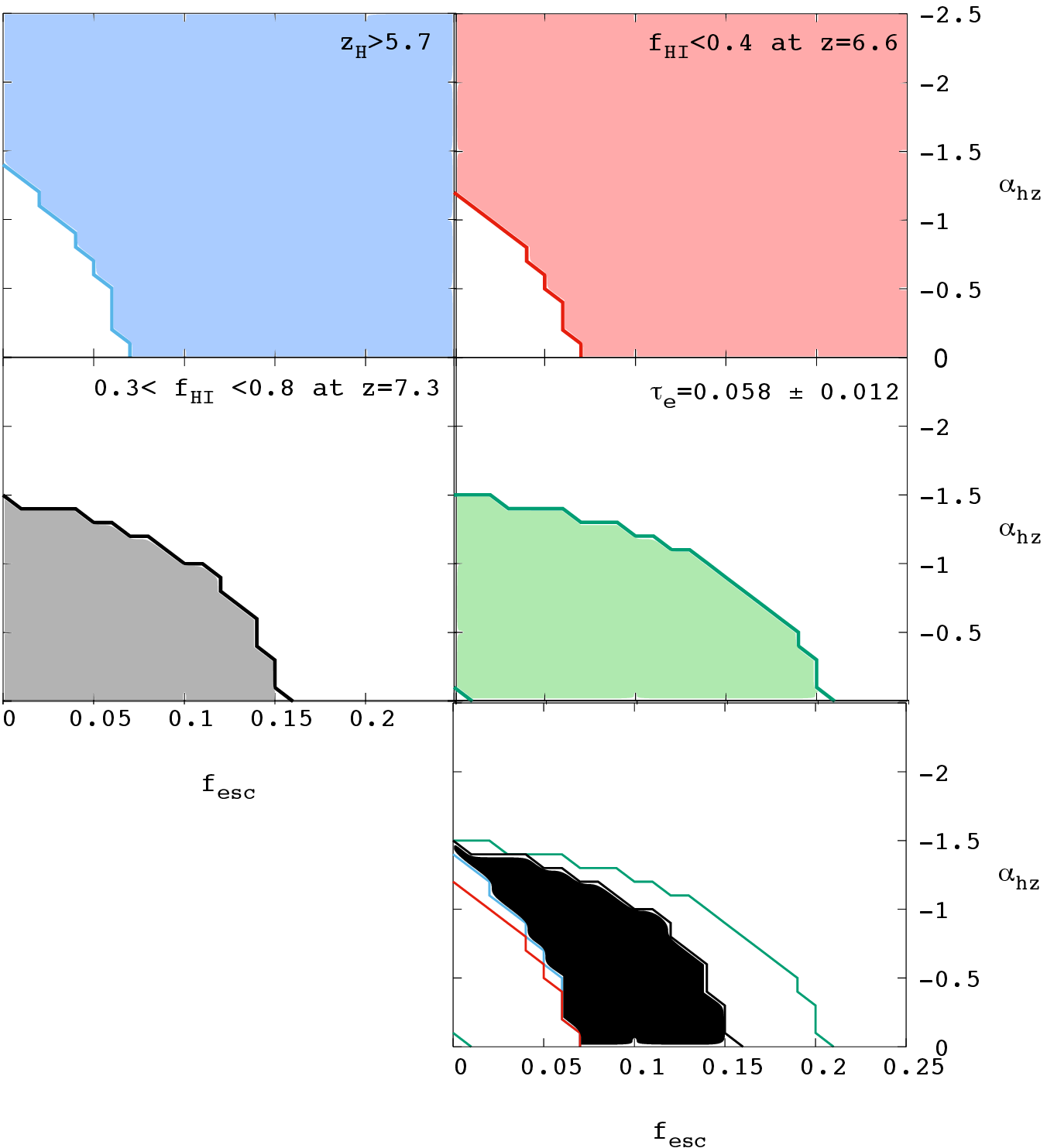}
	\caption{Observational constraints on $f_{\rm esc}$ and $\alpha_{\rm hz}$. 
	{\it Top left}: constraint imposed by the completion redshift of \ion{H}{i} reionization inferred from QSO spectra 
	$z_{\rm H} > 5.7$ \citep{2006AJ....132..117F}, 
	{\it Top right}: constraint imposed by the neutral fraction inferred from LAE LF 
	$f_{\rm \ion{H}{i}} < 0.4$ at $z=6.6$ \citep{2010ApJ...723..869O}. 
	{\it Middle left}: constraint imposed by the neutral fraction inferred from LAE LF 
	$0.3 < f_{\rm \ion{H}{i}} < 0.8$ at $z=7.3$ \citep{2014ApJ...797...16K}. 
	{\it Middle right}: the Thomson optical depth of the CMB $\tau_{\rm e} = 0.058 
	\pm 0.012$ \citep{2016arXiv160503507P}. 
	{\it Bottom right}: the combination of above four constraints. 
	The black filled region satisfy all of the constraints. }
\label{Params} 
\end{figure}

\begin{figure}
\centering
	\includegraphics[width=8.5cm]{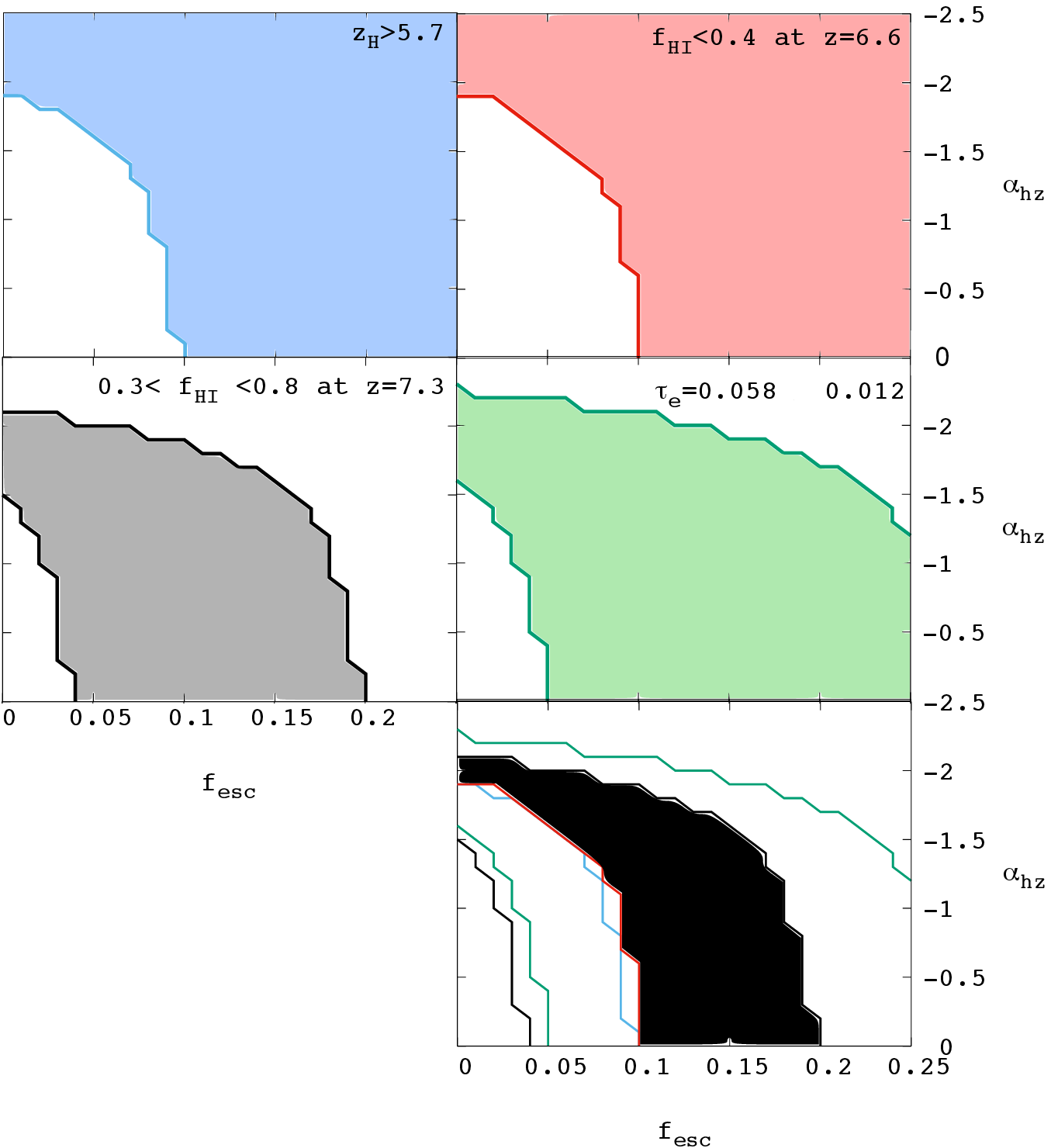}
	\caption{Same as Fig.~\ref{Params} but for PL model. 
	}
\label{ParamsPL} 
\end{figure}
Since the AGN abundance has a large uncertainty at $z>4$, hereafter we treat $\alpha_{\rm hz}$ as a free parameter. 
The escape fraction $f_{\rm esc}$ is also highly uncertain because direct measurement of $f_{\rm esc}$ is considerably difficult. 
Thus, we compare our results with the observational data of \ion{H}{i}/\ion{He}{ii} ionizing histories and Thomson optical depth of the CMB in order to obtain constraints on these parameters. 

\subsection{Dependence of the reionization history on $f_{\rm esc}$ and $\alpha_{\rm hz}$} 
Before the comparison with EoR observations, we first show the dependence of the reionization evolution on the parameters $f_{\rm esc}$ and $\alpha_{\rm hz}$.
Figs.~\ref{IH} and \ref{IHe} respectively show the \ion{H}{i} and \ion{He}{ii} reionization histories for different sets of $f_{\rm esc}$ and $\alpha_{\rm hz}$. 
In these figures, the reionization histories calculated with the AC and PL models are  represented by the thick and thin lines, respectively. 
We will describe the impact of the SED shape on the reionization history in \S~\ref{SEDshape}. 
As we naturally expect, the \ion{H}{i} reionization tends to be completed earlier as $f_{\rm esc}$ ($\alpha_{\rm hz}$) increases (decreases). 
While the \ion{He}{ii} reionization history has a stronger dependence on $\alpha_{\rm hz}$ than on $f_{\rm esc}$, because of the existence of X-ray photons that can ionize \ion{He}{ii} more effectively. 

The CMB optical depth to Thomson scattering is a good probe for the evolution of the EoR. The optical depth is obtained as \cite[e.g.] []{2016arXiv160503507P}
\begin{eqnarray}
	\tau_{\rm e}(z)=\int_0^{z} c \sigma_{\rm T} n_{\rm e} (1+z)^3 \frac{dt}{dz}dz ,
\end{eqnarray}
where $\sigma_{\rm T}$ is Thomson cross section. Fig.~\ref{tau} shows the dependence of $\tau_{\rm e}$ on the parameter set of $f_{\rm esc}$ and $\alpha_{\rm hz}$. 
The optical depth $\tau_{\rm e}$ does not change after the \ion{H}{i} reionization was completed, because almost all of electrons are produced by hydrogen ionization. Therefore, $\tau_e$ is sensitive to the hydrogen reionization history and thus to both of $f_{\rm esc}$ and $\alpha_{\rm hz}$

\subsection{Constraints on the model parameters}
There are several observations which provide constraints on the evolution of the \ion{H}{i} and \ion{He}{ii} fractions. 
From the comparison with our numerical results, we can obtain constraints on the parameter set of $f_{\rm esc}$ and $\alpha_{\rm hz}$. 
We summarize the constrained parameter set for the AC model in Fig.~\ref{Params}. 
The observations of the spectra of high-$z$ quasars suggest that the cosmic reionization of \ion{H}{i} is completed by $z=5.7$~\citep{2006AJ....132..117F}. 
Therefore, we set the criterion for the completion of \ion{H}{i} reionization as $f_{\rm \ion{H}{i}}<10^{-4}$ at $z_{\rm H}=5.7$. \cite{2010ApJ...723..869O} provided the constraint on the neutral fraction of $f_{\rm \ion{H}{i}}<0.2\pm0.2$ at $z=6.6$ from the observation of LAEs. Further, \cite{2014ApJ...797...16K} indicated $0.3<f_{\rm \ion{H}{i}}<0.8$ at $z=7.3$ from a drastic decrease in the number density of LAEs. 
Thus, the observations of LAEs and QSOs give the upper limit on the \ion{H}{i} fraction and set minimum amounts of ionizing sources. 
As we can see in top  panels of Fig.~\ref{Params}, they provide the lower (upper) bound on $f_{\rm esc}$ ($\alpha_{\rm hz}$). 
On the other hand, the lower bound on the \ion{H}{i} fraction at $z=7.3$ rules out the models with large values of $f_{\rm esc}$ and lower values of $\alpha_{\rm hz}$ as shown in the middle left panel of Fig.~\ref{Params}. 

Next we focus on the \ion{He}{ii} reionization. 
The effective optical depth for \ion{He}{ii} Lyman-$\alpha$ absorption suggests that almost all of helium in the IGM are fully ionized {at $z\sim2.7$} \citep{2001Sci...293.1112K,2004ApJ...600..570S,2010ApJ...722.1312S,2004ApJ...605..631Z,2006A&A...455...91F,2010ApJ...714..355F}.
However, with our LF model based on the evolution of $\Phi_*$ derived by \cite{2015A&A...578A..83G}, 
the AGN emissivity is sufficiently large to ionize \ion{He}{ii} by $z=2.7$ even if $\alpha_{\rm hz}=0.0$. 
Therefore, the observation of \ion{He}{ii} optical depth rule-out high emissivity AGN models (see also \cite{2016arXiv160602719M}).

The Planck observation~\citep{2016arXiv160503507P} measured the CMB optical depth for Thomson scattering and provided $\tau_{\rm e} = 0.058 \pm 0.012$. In Fig.~\ref{tau} we plot the lower and upper bounds on $\tau_{\rm e}$. Since the optical depth depends on $f_{\rm esc}$ and $\alpha_{\rm hz}$, we obtain the constraint on these parameters. We plot the upper and lower bounds on $f_{\rm esc}$ and $\alpha_{\rm hz}$ as shown in the bottom left panel of Fig.~\ref{Params}.

Combining all of the above constraints, we obtain the bottom right panel of Fig.~\ref{Params} which shows the parameter sets satisfying all constraints. 
Again, we emphasize that our model is based on the evolution of $\Phi_*$ derived from \cite{2015A&A...578A..83G}. 
As we can see, the evolution factor of the faint end slope of the AGN LF should be $\alpha_{\rm hz} < -1.5$ and the escape fraction must satisfy $f_{\rm esc} < 0.15$. 
The total constraint is mostly contributed from the observations of the \ion{H}{i} fraction at $z$ = 5.7 and 7.3. 
Employing the AC model, this result implies that high AGN abundance at high redshifts reported in \cite{2015A&A...578A..83G}, i.e., $\alpha_{\rm hz} \sim -1.5$  (see Fig.~\ref{Emis}), is not allowed unless the $f_{\rm esc}$ is considerably small. 
The allowed values of $f_{\rm esc}$ is relatively smaller than that predicted in the previous works, because the AGNs with AC model has non-negligible contribution to ionization as shown by Fig.~\ref{sed}. 

\subsection{Difference between the AC and PL models}\label{SEDshape}
In this section, we make the difference between the AC model and the PL model clear, and clarify how much the SED shape affects the constraints on $f_{\rm esc}$ and $\alpha_{\rm hz}$. 
We determine possible parameter sets for the PL model by the same way as the AC model, and summarize it in Fig.~\ref{ParamsPL}. 
Thin lines in Figs.~\ref{IH} to \ref{tau} represent the evolution of \ion{H}{ii}/\ion{He}{iii} fractions and $\tau_{\rm e}$ for the PL model with the same parameter sets as used for the AC model. 
First, in Fig. \ref{IH}, the whole \ion{H}{i} ionizing history is delayed if we employ the PL model. 
This indicates that, for the PL model, AGNs cannot ionize the \ion{H}{i} as effectively as the AC model. 
Then, the constraints imposed by \ion{H}{i} reionization and \ion{H}{i} fraction shift in the upper direction in the panels of Fig.~\ref{ParamsPL}. {In addition, the boundary curves become vertical for large values of $\alpha_{\rm hz}$. 
This result indicates that galaxies tend to be more responsible for the \ion{H}{i} reionization as $f_{\rm esc}$ increases. }

Second, similarly, the whole \ion{He}{ii} ionizing history is delayed. AGNs in the PL model emit less \ion{He}{ii} ionizing photons than the AC model. 
{However, as in the case of the AC model, the \ion{He}{ii} reionization does not reproduce the redshift of \ion{He}{ii} reionization at $z\sim2.7$. }

Third, the difference of $\tau_{\rm e}$ from the AC model reflects the delay in the ionizing history. 
Thus, the Thomson scattering optical depth $\tau_{\rm e}$ for the PL model is smaller than that for the AC model. 
Consequently, $\tau_{\rm e}$ provides not only the upper bounds but the lower bounds on the model parameters in the middle right panel of Fig.~\ref{ParamsPL}.

As a result, as we can see in the bottom right panel of Fig.~\ref{ParamsPL}, the combination of the aforementioned constrains the parameters as $f_{\rm esc} < 0.2$ and $\alpha_{\rm hz} > -2.0$. 
Besides, when the contribution from ANGs to reionization is relatively small, i.e. if $\alpha_{\rm hz} > -1.0$, $0.1< f_{\rm esc} < 0.2$ is required for satisfying all observational constraints. 
We mention that the required range of $f_{\rm esc} $ without constraints from $\Gamma_{\rm \ion{H}{i}}$ is almost consistent with the previous work \cite[e.g.,][]{2016MNRAS.457.4051K}. 
This indicates the AGNs for the PL model ionize \ion{H}{i} less effectively and therefore, a larger contribution from galaxies is required.
 
\subsection{Thermal history}
We can calculate not only the ionization history but also the thermal history of the IGM with our model. The ionizing history is affected by the thermal state of the ionized gas through the recombination coefficient. If we assume the fixed IGM temperature of 20,000~K as in MH15 for a model with $f_{\rm esc} = 0.10$ and $\alpha_{\rm hz} = -0.50$, $z_{\rm H}$ and $\tau_e$ increase by {0.12 and $1.0 \times 10^{-3}$} respectively. 

\begin{figure}
\centering
	\includegraphics[width=7cm]{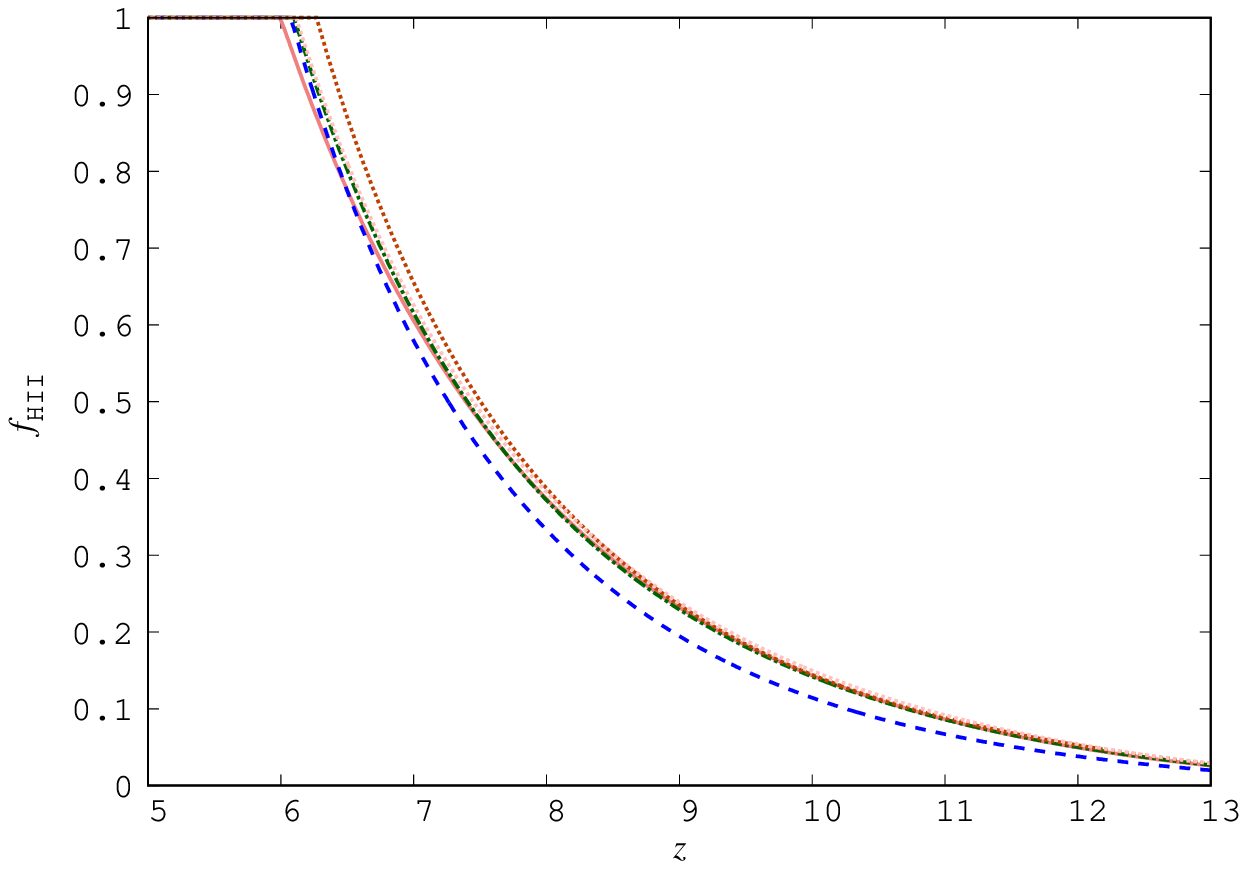}
	\includegraphics[width=7cm]{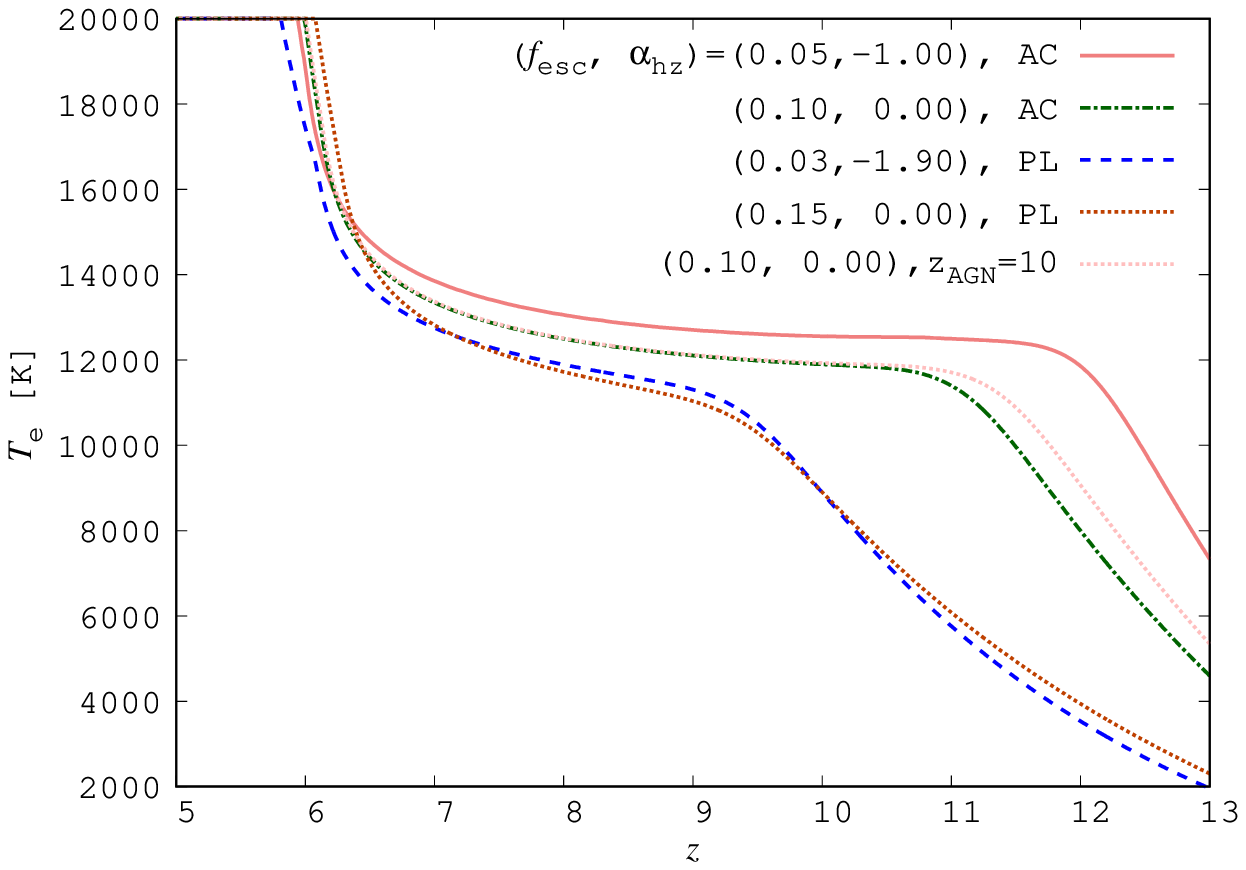}
	\caption{{\it Top}: evolution of the \ion{H}{ii} fraction for our constrained models with 
	some typical sets of the parameters. 
	The results obtained with ($f_{\rm esc}$, $\alpha_{\rm hz}$)~=~(0.05, -1.00) 
	and (0.10, 0.00) for the 
	AC model are represented by the solid and dot-dashed lines, respectively. 
	While the results obtained with ($f_{\rm esc}$, $\alpha_{\rm hz}
	$)~=~(0.03, -1.90) and (0.15, 0.00) for the PL model are shown by the dashed and dotted 
	lines, respectively. 
	The result with ($f_{\rm esc}$, $\alpha_{\rm hz}
	$)~=~(0.10, 0.00) and $z_{\rm AGN}=10$ are also shown by the thin dotted line. 
	{\it Bottom}: thermal history for the same models. 
	All of the reionization histories are almost the same, but the thermal histories 
	are obviously different depending on the abundance of AGNs at high redshifts and 
	the SED type of AGNs. 
	The thermal history is more sensitive to the SED type. } 
\label{Thermal} 
\end{figure}

Besides, in Fig.~\ref{Thermal}, we show the \ion{H}{i} reionization and thermal histories computed with two representative parameter sets, i.e., ($f_{\rm esc} $, $\alpha_{\rm hz}$)~=~(0.05,~-1.00)  and (0.10,~0.00), for the AC model. 
Although these two reionization histories are almost identical, the IGM temperature in the former case (solid line) is higher than that in the latter (dot-dashed line), because the heating efficiency of AGNs is much higher than galaxies and thus more abundant AGNs results in higher IGM temperature. 
Therefore, the thermal history could be another probe of ionization sources.

{Furthermore, we show the reionization and thermal histories computed with the PL model. 
The parameters $f_{\rm esc}$ and $\alpha_{\rm hz}$ are chosen so that similar evolution of $f_{\rm \ion{H}{ii}}$ is realized. 
Since AGNs in the PL model radiate relatively smaller amount of high-energy photons compared to the AC models for a fixed amount of ionizing photons, it turns out that employing the PL model results in clearly different thermal history. 
The difference in the IGM temperature can be an indicator of the existence of AGNs at high redshifts and a probe of the SED shape. 
Here note that the thermal history is much more sensitive to the SED shape than $\alpha_{\rm hz}$. }
Increasing $z_{\rm AGN}$ also provides more efficient heating at higher redshifts.  
The IGM temperature in the case of $z_{\rm AGN}=10$ is higher than that of the fiducial model with the same values of $f_{\rm esc}$ and $\alpha_{\rm hz}$ because the larger value of $z_{\rm AGN}$ leads to high AGN emissivity at higher redshifts. 

The thermal history is expected to be imprinted on 21cm-line signals from neutral hydrogen through the spin temperature. The 21cm-line signals are a major target of the SKA and its pathfinders such as the MWA and LOFAR. In particular, statistical quantities such as the power spectrum, bispectrum, variance and skewness are expected to be able to probe the fluctuation in the 21cm brightness temperature \citep{1997ApJ...475..429M,2000ApJ...528..597T,2006PhR...433..181F,2007MNRAS.376.1680P,2014ApJ...782...66P,2010A&A...523A...4B,2015MNRAS.451..467S,2015MNRAS.451..266Y}.

\section{Discussion}\label{discussion}

\subsection{Model uncertainty}

In this subsection, we discuss the uncertainties in our model.
{First, let us consider the SED and LF of AGNs. We assumed that the SED depends on the Eddington ratio $\lambda_{\rm Edd}$ and all AGNs have the same value for a fixed redshift. 
In spite of the fact that some previous studies reported dependences of $\lambda_{\rm Edd}$ on AGN luminosity \cite[e.g.,][]{2004A&A...426..797C,2012ApJ...761..143N}, we ignored the dependence since it has large variance. 
For example, when faint (bright) AGNs have relatively small (large) values of $\lambda_{\rm Edd}$, their ionization efficiency would become smaller (larger). Then, the constraint on $\alpha_{\rm hz}$ will become weaker if we take this effect into account. 
{Besides, we need to mention that there are other studies reporting less abundant AGNs at $z>3$ than \cite{2015A&A...578A..83G}  \cite[e.g.][]{2015MNRAS.453.1946G,2015MNRAS.448.3167W}. 
If we employ luminosity functions obtained in their study, the allowed parameter regions in Figs.~\ref{Params} and \ref{ParamsPL} systematically shift towards the upper right-hand side.}
}

Next, concerning star forming galaxies, we used a fixed SFRD model. However, the SFRD depends on the faint-end shape and dust correction of UV luminosity functions. Although some recent observation indicated the observational faint-end limit of UV magnitude $M_{\rm UV}=-17$ and the faint-end slope $\alpha=-2$ at $z>6$ \citep{2015ApJ...799...12I,2015ApJ...803...34B}, UV luminosity functions during the EoR are still less known. For instance, \cite{2015ApJ...799...12I} showed that the SFRD integrated down to $M_{\rm UV}=-10$ is larger than one down to $M_{\rm UV}=-17$ by a factor of 3, assuming $\alpha=-2$. Our SFRD model \citep{2014ARA&A..52..415M} was obtained by integration of luminosity function down to $0.03L^{*}$, where $L^{*}$ is the characteristic luminosity of the Schechter function. Further, the dust correction causes a variation by a factor of 2 to the SFRD \citep{2006ApJ...651..142H}. 
Additionally, a rapid decrease in the SFRD at $z>8$ is indicated from a recent observation \citep{2015ApJ...803...34B}. Thus, the contribution of galaxies to the \ion{H}{i} reionization at $z>8$ may be overestimated in our calculations. 
For these reasons, the SFRD has an uncertainty of a factor of 2 $\sim$ 6. 
In our calculation, $f_{\rm esc}$ should be considered to incorprate this uncertainty.

\begin{figure}
\centering
\includegraphics[width=7cm]{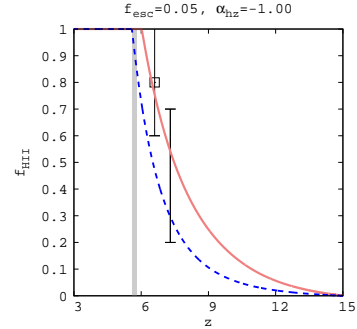}
\caption{\ion{H}{i} ionization history homogeneous (solid line) and inhomogeneous (dashed line) IGM.}
\label{teHI} 
\end{figure}

Finally, let us see the effect of the inhomogeneity of the IGM. 
We treated the IGM as being uniform in our calculations, however the observed Universe is obviously inhomogeneous.  
The effective optical depth $\tau_{\rm eff}$ through the inhomogeneous medium is estimated as an integration of the \ion{H}{i} distribution function and the continuum optical depth with respect to \ion{H}{i} column density \citep{2012ApJ...746..125H}. 
In \cite{Inoue 2014}, they reported the distribution function of intergalactic absorbers (IGA) $F_{\rm IGA}(z)$ = $\partial^2 n/\partial z\partial N_{\rm \ion{H}{i}}$ at $z<6$ and gave an analytic model. 
\cite{2016MNRAS.457.4051K} have shown that an \ion{H}{i} distribution function derived from a cosmological hydrodynamic simulation well corresponds to the analytic model by \cite{Inoue 2014} at $z=5-6$, and used it for computing the reionization history. 
Despite the efforts by these studies, there is still large uncertainty in the \ion{H}{i} distribution function during the EoR; we should estimate it to be consistent with the reionization process that affects the \ion{H}{i} distribution function. 
Since constructing the complete model is very tough work, we evaluate the impact of the inhomogeneity with a simple model. 
We adopt the analytic model by \cite{Inoue 2014} in post-reionization epoch and extrapolate to higher redshifts as $F_{\rm IGA}(z) = F_{\rm IGA}(z=5.7) \left(f_{\rm \ion{H}{i}}(1+f_{\rm \ion{H}{ii}}^{\rm ion})/f_{\rm \ion{H}{ii}}^{\rm ion}(1-f_{\rm \ion{H}{ii}}^{\rm ion})\right) \left((1+z)/(1+5.7)\right)^2$. Here, $f_{\rm \ion{H}{ii}}^{\rm ion} = 3 \times10^{-3}$ corresponds to the mass weighted ionized fraction at $z=5.7$ \citep{2006AJ....132..117F} . For simplicity, we assume that the distribution of \ion{He}{i} and \ion{He}{ii} follows the \ion{H}{i} distribution. Therefore, we estimate the column density of \ion{He}{i} and \ion{He}{ii} as $N_{\rm \ion{He}{i}}=(f_{\rm \ion{He}{i}}/({f_{\rm \ion{H}{i}}}+f_{\rm \ion{H}{ii}}^{\rm ion}))~(n_{\rm He}/n_{\rm H})~ N_{\rm \ion{H}{i}}$ and $N_{\rm \ion{He}{ii}}=(f_{\rm \ion{He}{ii}}/({f_{\rm \ion{H}{i}}}+f_{\rm \ion{H}{ii}}^{\rm ion}))~(n_{\rm He}/n_{\rm H})~N_{\rm \ion{H}{i}}$. As a result, as we can see in Fig.~\ref{teHI}, a model with $f_{\rm esc}=0.05$ and $\alpha_{\rm hz}=-1.00$ completes \ion{H}{i} reionization at $z_{\rm H}=5.58$ in clumpy universe, while the completion is at $z_{\rm H}=6.02$ for uniform IGM. Therefore, the appropriate value of $f_{\rm esc}$ ($\alpha_{\rm hz}$) may become larger (smaller) than our results if we incorporate the inhomogeneity of the IGM. 

\begin{figure*}
\centering
\includegraphics[width=13.5cm]{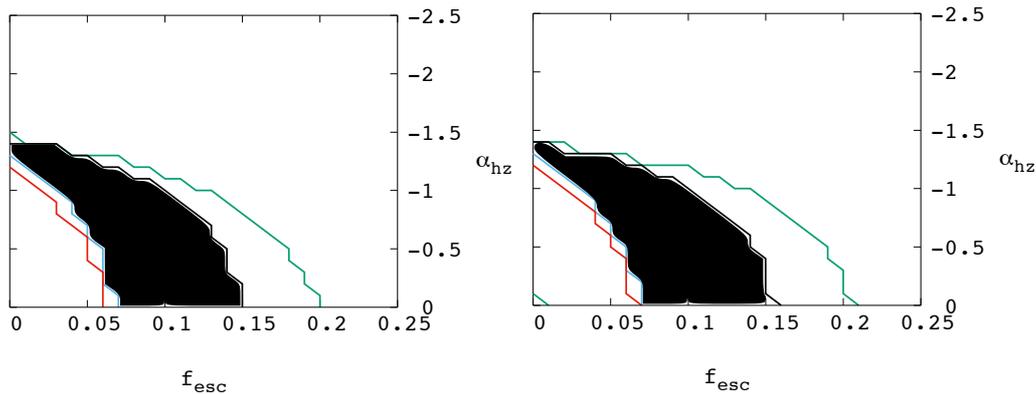}
\caption{Constraints on the parameters, $f_{\rm esc}$ and $\alpha_{\rm hz}$ for two kind of models with $z_{\rm AGN}=10$ (left) and $M_{\rm min}=-15$ (right). The black regions are combined allowed regions. }
\label{CompResults} 
\end{figure*}

\subsection{Remaining parameters} 
In this work, we mainly focused on $f_{\rm esc}$ and $\alpha_{\rm hz}$. 
However, our model has two other parameters, $z_{\rm AGN}$ and the faint end of UV luminosity $M_{\rm UV}^{\rm min}$. 

Firstly, we discuss the $z_{\rm AGN}$ which determines the critical redshift of AGN abundance decrease at higher redshifts. This parameter relates to many interesting questions on the origin of the first AGNs. The value of $z_{\rm AGN}$ would be larger than $6$ where AGNs have been observed, but it is highly uncertain. When we set $z_{\rm AGN}=10$ rather than our fiducial value $z_{\rm AGN}=6$, AGNs ionize \ion{H}{i} at higher redshifts. In the left panel of Fig.\ref{CompResults}, we show the allowed parameter region for calculations with $z_{\rm AGN}=10$. Consequently, we find that the result is not different significantly with the fiducial one.

Secondly, we discuss the faint end of the AGN UV luminosity function. Although a faint AGN with $M_{\rm UV}=-19$ has been found at $z=5.75$ in \cite{2015A&A...578A..83G}, the faint end has not been well determined observationally even at lower redshifts. However, we did not treat $M_{\rm UV}^{\rm min}$ as a main parameter in this work, because faint AGNs (-18 < $M_{\rm UV} < -15$) do not significantly contribute to the total AGN luminosity as far as we consider the faint-end slope of $\alpha_{\rm hz} > - 2.0$ (the bottom panel of Fig.~\ref{sed}). 
Therefore, the results are expected to be rather insensitive to the value of the faint end. This can be confirmed in the right panel of Fig.\ref{CompResults} where the faint end is set to be  $M_{\rm UV}^{\rm min} = -15$. 

\subsection{HI Photo-ionization Rate}
{As we mentioned in \S~2.2.3, extreme abundant AGN model provides more $\rm \ion{H}{i}$ photo-ionization rate than observationally measured values. 
There are several measurements at $z\sim2-6$ such as the Lyman-$\alpha$ opacity of the IGM \citep{2013MNRAS.436.1023B} and the proximity effect \citep{2008A&A...491..465D,2009arXiv0906.1484D}. Here we set a conservative upper limit, $3\times10^{-12}$, at $z\sim2-4$ because these observed values have a large uncertainty and conflicts with each other (e.g. figure.~13 in \cite{Calverley11}). Then, we evaluate the photo-ionization rate of AGNs as,
\begin{eqnarray}
\Gamma_{\rm \ion{H}{i}}^{\rm AGN} = \int_{\nu_{\rm \ion{H}{i}}}d\nu\frac{F(z,\nu)}{h_{\rm p}\nu}\sigma_{\rm \ion{H}{i}}(\nu) [\rm s^{-1}]. 
\end{eqnarray}
Thus, we find that this constraint provides a lower limit of $\alpha_{\rm hz}>-1.6$ for AC model and $\alpha_{\rm hz}>-1.9$ for PL model. This additional constraint hardly changes the lower limit of $\alpha_{\rm hz}$ for the AC model. However, for PL model, the constraint rules out the AGN-driven reionization. Despite these additional constraints, the existence of abundant faint AGNs at high redshifts as suggested in \cite{2015A&A...578A..83G} is allowed.
}

\begin{figure}
	\centering
	\includegraphics[width=6.5cm]{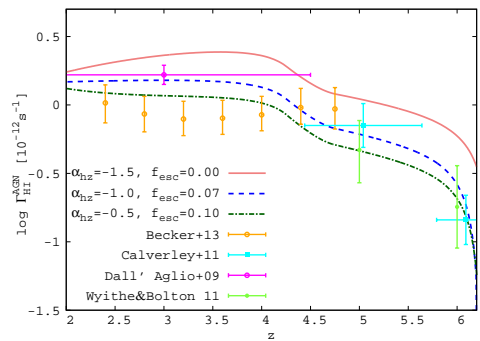}
	\caption{{\ion{H}{i} photo-ionization rate of AGN for three kinds of parameter sets with AC model. The results with ($\alpha_{\rm hz}$, $f_{\rm esc}$)=(-1.5, 0.0), (-1.0, 0.07) and (-0.05, 0.10) are shown by red solid line, blue dashed line and green dot-dashed line. It should be note that \ion{H}{i} photo-ionization rate of galaxies are not included in these lines. We plot \ion{H}{i} photo-ionization rate inferred from literatures with 1$\sigma$ error : \citep{2013MNRAS.436.1023B} in yellow, \citep{Calverley11} in cyan, \citep{2009arXiv0906.1484D} in magenta and \citep{2011MNRAS.412.1926W} in green.}}
\label{PhotoIon} 
\end{figure}

{
In Fig.~\ref{PhotoIon}, we show the redshift evolution of $\Gamma_{\rm \ion{H}{i}}^{\rm AGN}$ obtained from AC models with three different parameter sets satisfying all observational constraints in Fig.~\ref{Params}. 
We also plot $\Gamma_{\rm \ion{H}{i}}^{\rm AGN}$ obtained from observations for comparison \citep{2013MNRAS.436.1023B, Calverley11,2009arXiv0906.1484D,2011MNRAS.412.1926W}. The three lines of our models are consistent with constraints in the range of 3$\sigma$ of constraints. 
}

{
We note that $f_{\rm \ion{H}{i}}$ is artificially fixed after reionization in our analysis. However, in practice, it evolves at $z<6$ and the photo-ionization rate has a variation proportional to $f_{\rm HI}$. In addition, we do not consider the contribution to \ion{H}{i} photo-ionization rate from galaxies. {Thus, the \ion{H}{i} photo-ionization rate is increased when including the galactic contribution.} According to \cite{2016MNRAS.457.4051K}, the \ion{H}{i} photo-ionization at $2<z<4$ requires $f_{\rm esc}<0.1$ which conflicts with our constraints except for an abundant AGN model with $\alpha_{\rm hz}<-1$. Thus, galaxy-dominant reionization model needs a decrease in $f_{\rm esc}$ at $z<4$. 
}

\subsection{Cosmic X-ray background}
{As we showed, high-redshift AGNs can significantly contribute to the reionization of intergalactic hydrogen and helium. Then there is a possibility that they contribute to unresolved X-ray background (XRB) as well. 
In this subsection, we estimate the unresolved XRB coming from AGNs and discuss whether the unresolved XRB can place a additional constraint on the parameter of AGN abundance
}

\begin{figure}
\centering
	\includegraphics[width=6.5cm]{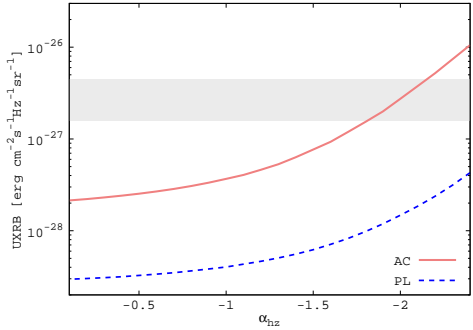}
	\caption{{Unresolved X-ray background (XRB) spectrum at 2keV for 
	the AC model 	(solid line) and the PL model (dashed line) as a function of 
	$\alpha_{\rm hz}$. 
	The unresolved XRB derived by \citet{2012A&A...548A..87M} is indicated by 
	the shaded region. 
	For the AC model, the numerically computed 
	unresolved XRB exceeds the observed value if faint AGNs are extremely abundant 
	as much as $\alpha_{\rm hz}<-2$, although this constraint does not dominate
	the other constraints shown in Fig. ~\ref{Params}. }}
\label{XRB} 
\end{figure}

We simply assume the flux from AGNs at $z>5$ to be the unresolved XRB at the preset epoch. 
It should be noted that the intensity includes the bright AGNs and, in practice, there is contribution from faint ANGs at $z<5$. 
In Fig.~\ref{XRB}, we show the dependence of unresolved XRB at 2~keV on $\alpha_{\rm hz}$ for both the AC and PL models. 
The unresolved XRB at 2~keV is estimated as 
\begin{eqnarray}
J[{\rm 2keV}] = \frac{F(z=0,\nu_{\rm 2keV})}{4\pi}.
\end{eqnarray}
By comparing observationally proposed value $J[\rm2keV]=2.9^{+1.6}_{-1.3}\times10^{-27}[\rm erg~cm^{-2}~s^{-1}~Hz^{-1}sr^{-1}]$ \citep{2012A&A...548A..87M}, we find that $\alpha_{\rm hz}<-2.1$ is prohibited for the AC model and hence the unresolved XRB does not work as an additional constraint on $\alpha_{\rm hz}$ (see Fig.~\ref{Params}). 

\subsection{Radio loud quasars}

If AGNs were too abundant and bright in radio frequency range in high
redshift, their emission would serve as a background of the 21cm
line of neutral hydrogens. In this subsection we estimate, using our constrained AGN model, the number of photons that radio loud QSOs would emit at radio frequency range and compare it to that of the CMB.  We base our estimation on the result of \cite{1994ApJS...95....1E,2010MNRAS.402..724P}, in which the authors found SED of radio loud QSOs at radio frequency range to be 2-3 orders of magnitude smaller than the SED at $E \approx10$~eV. 
Remembering that the intensity ratio between at $1.25\mu$m and $1$keV is approximately $10^{-0.4}$, it is expected that the intensity ratio between at $21$ cm and $1$ keV is about $\sim 10^{-2}$. Setting the ratio as $10^{-\alpha}$ ($\alpha=2$ is the fiducial value), we write the SED at $21$ cm line frequency as
\begin{eqnarray}
\frac{F_{\nu_{21}}}{4\pi}
&=& \frac{10^{\alpha}}{4\pi}
    \left(\frac{\nu_{1\rm keV}}{\nu_{21 \rm cm}}\right)F_{1\rm keV}
\end{eqnarray}
The conversion formula from intensity to brightness temperature is given by
\begin{eqnarray}
\left(\frac{T_b}{\rm K}\right)
= 3.255 \times 10^{-5} \left(\frac{\nu}{\rm GHz}\right)^{2}
  \left(\frac{F_{\nu}/4\pi}{\rm Jy~str^{-1}}\right)~.
\end{eqnarray}
We evaluate as $F_{\nu_{1\rm keV}}/4\pi =1.79\times10^{-3}$ $\rm Jy/str$ at $z=6$ for AC model with $\alpha_{\rm hz}=-1.5$, and $T_b < 10 \times 10^{-\alpha}$[K] in terms of brightness temperature. If we assume that the AGNs are all radio loud QSOs ($\alpha=2$), this estimate gives $T_b < 0.1$~K. Therefore, the contribution of radio loud AGNs to the background of 21cm line should be negligible compared to the CMB $(T_\gamma = 2.7(1+z))$. 

\section{Conclusion}

In this work, we studied the \ion{H}{i} reionization with a simple model of star forming galaxies and AGNs. {Then, we employed theoretical motivated SED model and power law SED model. } We compared the observational data on the \ion{H}{i} fraction and CMB optical depth with our model and obtained constraints on the escape fraction of ionizing photons, $f_{\rm esc}$, and the faint-end slope of the AGN luminosity function at $z>4.25$, $\alpha_{\rm hz}$. As a result, we have found that models with $ \alpha_{\rm hz} > -1.5$ and $f_{\rm esc} < 0.15$ are consistent with the existing observations and the constraints come mostly from the \ion{H}{i} fraction history. Our result suggests that an AGN-dominated model with AGN abundance as large as the estimation in \cite{2015A&A...578A..83G} is allowed, while a galaxy-dominated model is also allowed. 
{It should be noted that we model the redshift evolution of AGN luminosity function so as to let $\rm M_*$ and $\Phi_*$ be consistent with those in \cite{2015A&A...578A..83G}. 
This keeps AGNs to ionize the Universe even with $\alpha_{\rm hz}=0$ and requires lower $f_{\rm esc}$.}
We have also calculated the thermal evolution of the IGM at $z>6$ and found the thermal evolution is sensitive to the abundance of AGNs and the shape of the SED. Therefore, the forthcoming 21cm-line observation will be useful to obtain more powerful constraints on properties, i.e. abundance and SED, of ionizing sources. {Recently, \cite{2017arXiv170104408K} have shown that the AGN dominant model enhances the 21cm power spectrum on large scales. }

{We have discussed that the \ion{H}{i} photo-ionization rate of AGNs and the contribution from AGNs to the unresolved XRB, and found that extremely abundant AGN models conflict with the observed \ion{H}{i} photo-ionization rate and unresolved XRB measurements. }
Furthermore, even if we assume that the radio loud AGNs exist at high redshifts, the radiation is too small to serve as background emission of the 21cm-line signal of neutral hydrogen. 

\section*{Acknowledgement}
We are grateful to A. K. Inoue for his suggestions regarding the model of AGNs, to T. Kawaguchi for kindly providing us with the theoretical and observational SED data, and to H. Shirakata for providing simulation results. We also thank the members of SKA-JP EoR science woking group for fruitful discussion and comments. 
This work is supported by Grant-in-Aid from the Ministry of Education, Culture, Sports, Science and Technology (MEXT) of Japan, No. 26610048, No. 15H05896, No 16H05999 (K.T.), 24340048 (K.T., K.I.), No. 25-3015 (H.S.), 15K17646 (H.T.), 16J01585 (S.Y.). and a grant from NAOJ. H.T. also acknowledges the support by MEXT Program for Leading Graduate Schools PhD professional, ``Gateway to Success in Frontier Asia''.




%
%


\bsp	
\label{lastpage}
\end{document}